\documentclass[12pt,preprint]{aastex}
\newcommand{\blank}[1]{}

\newcommand{\chg}[1]{{\bf #1}}
\slugcomment{draft version}

\begin{document}

\title{Spectral Statistics and Local Luminosity Function of a Hard X-ray Complete Sample of Brightest AGNs}

\author{K. Shinozaki}
\affil{Department of Physics, Tokyo Metropolitan University,
    1-1 Minami Osawa, Hachioji, Tokyo 192-0397}
\email{shino@phys.metro-u.ac.jp}
\author{T. Miyaji}
\affil{Department of Physics, Carnegie Mellon University,
Pittsburgh, PA 15213, USA}
\email{miyaji@cmu.edu}
\author{Y. Ishisaki}
\affil{Department of Physics, Tokyo Metropolitan University,
    1-1 Minami Osawa, Hachioji, Tokyo 192-0397}
\email{ishisaki@phys.metro-u.ac.jp}
\author{Y. Ueda}
\affil{ISAS/JAXA, 3-1-1 Yoshinodai, Sagamihara, Kanagawa 229-8510}
\email{ueda@astro.isas.ac.jp}
\author{and}
\author{Y. Ogasaka}
\affil{Department of Astrophysics, Nagoya University, Nagoya, Aichi 464-8602}
\email{ogasaka@u.phys.nagoya-u.ac.jp}
%\author{K. Hayashida}
%\affil{Department of Astrophysics, Osaka University, Matikane, Toyonaka,
%Osaka, 560-0043}
%\email{hayasida@ess.sci.osaka-u.ac.jp}
%\author{H. Awaki}
%\affil{Department of Physics, Ehime University, Matsuyama, Ehime 790-8577}
%\email{awaki@sgr.phys.sci.ehime-u.ac.jp}
%\and
%\author{G. Hasinger}
%\affil{Max-Planck-Institut f\"ur extraterrestrische Physik, 
%Postf. 1312, D-85741 Garching, Germany}

%%% TMEDIT

\begin{abstract}

We have measured the X-ray spectral properties
of a complete flux-limited sample of bright AGNs from {\it HEAO-1} 
all-sky catalogs to investigate their statistics and 
provide greater constraints on the bright-end of the 
hard X-ray luminosity function (HXLF) of AGNs and
the AGN population synthesis model of the X-ray background.
Spectral studies using data from {\it ASCA}, {\it XMM-Newton}
and/or {\it Beppo-SAX} observations have been made  
for almost all AGNs in this sample.

The spectral measurements enable us to construct 
the neutral absorbing column density ($\log\,N_{\rm H}$)
distribution and separate HXLFs
for absorbed ($\log\,N_{\rm H}[{\rm cm^{-2}}]> 21.5$) and 
unabsorbed AGNs in the local universe.
Our results show evidence for a difference in the shapes of HXLFs of
absorbed and unabsorbed AGNs in that absorbed AGN HXLF drops
more rapidly at higher luminosities than that of unabsorbed AGNs,
which is similar to that previously reported.
In the $L_{\rm X} - N_{\rm H}$ plot, we found no AGN in the 
high-luminosity high-intrinsic absorption regime 
($\log\,L_{\rm X}[\rm erg/s]> 44.5$,  
$\log\,N_{\rm H}[\rm cm^{-2}]> 21.5$) in our sample,
where we expect $\sim$ 5 AGNs if we assume that  
absorbed and unabsorbed having identical AGN HXLF shapes.
We also find that the observed flux with {\it ASCA} or 
{\it XMM-Newton} is smaller than that with {\it HEAO-1} by a 
factor of 0.29 on average, which is expected for re-observation 
of sources with a factor $\sim$ 2.5 variability amplitude scale.

\end{abstract}


\section{INTRODUCTION}

% AGNs (Active Galactic Nuclei) 
% it suggest that AGNs have super massive black hole in the center.
Investigating the X-ray luminosity function (XLF) of AGNs in different 
redshifts is very important in understanding the growth of super massive
black holes (SMBH) at the centers of galaxies over the cosmological 
time scale, which is directly linked to the accretion history of 
the universe.
%
% Update these statements 
%
 Recent progress on X-ray surveys has enabled us to trace the 
evolution of the XLF over a wide redshift and 
luminosity ranges. In particular, in the soft X-ray ($0.5-2.0$~keV)
band, a combination of extensive {\it ROSAT} surveys 
\citep{Miyaji2000a} has been extended by recent {\it Chandra} and 
{\it XMM-Newton} surveys \citep{Hasinger2005}. With this  
combination,consisting of $\sim 1000$ type 1 AGNs, they traced 
the evolution of the soft X-ray luminosity function (SXLF) with 
a high precision. In particular, they clearly traced the 
so-called anti-hierarchical AGN evolution (some call 'down-sizing'), 
where the number density peak of lower luminosity
AGNs comes much later in the history of the universe ($z\sim 1$)
than that of higher luminosity ones ($z\ga 2$). A fundamental 
limitation of the soft X-ray sample, however, is that it is not 
sensitive to type 2 AGNs, in which AGN activities are heavily 
obscured by photoelectric absorption by intervening gas,
except those at very high redshifts.

 Hard X-ray ($2<E<7-10$~keV) surveys are sensitive also to 
type 2 AGNs, up to a column density of 
$N_{\rm H}\sim 10^{24}$ cm$^{-2}$ (Compton-thin), giving a much 
more direct measure of the accretion history onto SMBH. Several 
groups constructed hard X-ray luminosity functions (HXLFs) of 
AGNs in this band 
\citep{Ueda2003,Barger2005,Silverman2005,Lafranca2005}. 
Hard X-ray surveys are limited in the number of 
AGNs as well as completeness, partially due to lower sensitivity of the
instruments to harder X-rays and also due to the difficulty of
identifying optically faint type 2 AGNs. In spite of these difficulties,
these HXLF studies came to a point showing the anti-hierarchical
evolution trend.

The XLF and X-ray spectra of the constituent AGNs are key elements of the
``AGN population synthesis modeling'', initially aimed at
explaining the Cosmic X-ray Background (CXRB) \citep[e.g.][]
{Madau1994,Comastri1995,Gilli2001}, where the contributions
of cosmologically evolving populations of AGNs with different absorption 
columns ($N_{\rm H}$) are considered to synthesize the CXRB spectrum, 
source number counts and the distribution of column densities in 
X-ray surveys with various depths, areas and energy bands. 
In order to test/constrain
these models, a complete samples with well defined flux limits as a 
function of survey area is essential. Investigating the spectral 
properties such complete samples models will be particularly strong
observational constraints by providing, e.g. the distribution of 
$N_{\rm H}$.   \citet{Ueda2003} (hearafter U03)
constructed an HXLF in the $2-10$~keV luminosity range of
$\log L_{\rm x}$ (erg s$^{-1}$) $\sim 41.5 - 46.5$ as a function 
of redshift up to 3, using an extensive set of highly complete 
samples ($>$ 96$\%$ total) of 247 AGNs from Hard X-ray surveys from
{\it HEAO-1}, {\it ASCA}, {\it CHANDRA}. U03 also
integrated the X-ray spectral (or hardness) information,
with the intrinsic absorption column density as a main spectral
parameter  into the HXLF to construct a population synthesis model.
One of the important findings of U03's analysis is that the ratio 
of obscured to unobscured AGNs decreases with luminosity, as 
(less quantitatively) suggested by \citet{Lawrence1982} and 
\citet{Miyaji2000b}.
This has been verified in the nearby AGNs by \citet{Sazonov2004} 
from a sample of 95 AGNs detected in the $3-20$~keV band during the 
{\it RXTE} slew survey \citep{Revnivtsev2004}. 

%%$B$b$7$/$O$I$3$+$K$3$l$b$$$l$k(B
%%it had a implication that the fraction of X-ray absorbed AGNs decreases
%%with the intrinsic luminosity. 

%%But until now, HEAO-1 All Sky Survey data is one of the only catalog 
%%with the large area survey in Hard X-ray band, 
%%and a study of AGN statistics used to have a
%%incomplete sample with roughly information of the spectrum of each samples,
%%it indicates that we could not investigate the 
%%population study of AGNs with the selection bias correctly of the
%%sample detection.

 In order to investigate the nature and evolution of the X-ray emission 
of AGNs, statistical characteristics of the X-ray sources in the 
bright-end sample (mostly consisting of present-day population) 
from large area surveys is important to compare with the fainter (high-$z$) 
population from deep surveys. The sample has to be 
complete with well-defined criteria, so that we can adequately account 
for selection biases. The spectral properties of all sources of the 
sample should be analyzed with good spectral resolution data
(for example, about 1000~cts with 7~ks of XMM-Newton / 
1000~cts with 20~ks 
of ASCA for 5e-12~cgs) to constrain $\Gamma$ to $\pm 0.03$ and 
$\log N_{\rm H}$ to $\pm 0.3$.
We also note that there are many absorbed AGNs, the accurate measurements
of $N_{\rm H}$ enables us to calculate de-absorbed luminosity, 
which is a more direct indicator of the intrinsic AGN power and thus 
the mass accretion rate.
One such study has been made by \citet{Schartel1997}  for 
\citet{Piccinotti1982} sample from {\it HEAO-1} A2 using {\it ROSAT} and
{\it EXOSAT} as the source of spectroscopic information. 
A recent study of {\it RXTE} Slew Survey catalog \citep{Revnivtsev2004}
utilized by \citet{Sazonov2004} has absorption estimates based on hardness 
ratio in two {\it RXTE} bands, but this method is only sensitive to 
neutral absorbing
column densities of $\log N_{\rm H}$ (cm$^{-2}$) $>22$. 
In deeper regimes, \citet{Mainieri2005,Mateos2005a,Mateos2005b} made 
extensive spectral analysis on the X-ray sources detected in a 
{\it XMM-Newton} serendipitous and deep 
Lockman Hole surveys, sampling intermediate to high redshifts.
In view of these, we have 
compiled the results of high-quality intermediate-resolution spectroscopic 
analysis for a complete hard X-ray flux-limited sample of AGNs
defined from {\it HEAO-1} all-sky survey catalogs. Detailed spectral 
information of almost all of the AGNs in this sample have been obtained by  
making spectral analysis and/or by referring to the spectral fit results from 
literature using data from {\it XMM-Newton}, {\it ASCA}, and {\it Beppo-SAX}.
The results of our preliminary analysis has been integrated into the global 
HXLF analysis over cosmological timescales by U03. In this paper, 
we present full results of our spectral analysis with necessary 
refinements and updates.

 The scope of this paper is as follows. In Sect.~\ref{sec:Sample}, we 
explain the construction of our sample. In Sect.~\ref{sec:Analysis}, 
we explain the spectral analysis of the {\it XMM-Newton} and {\it ASCA}
data. The statistical properties of the spectral results, including the 
distributions of spectral parameters are presented in  
Sect.~\ref{sec:Resultspectrum}. 
In Sect.~\ref{sec:HXLF},
we construct the local HXLFs separately for absorbed and unabsorbed 
AGNs from our sample and show the decrease of the fraction of
absorbed AGNs towards high luminosities. The overall results are discussed 
in Sect.~\ref{sec:Disc}. In Appendix~\ref{sec:Conf}, we explain our 
first-order correction to count rate and the effective survey area function
for a bias due to confusion noise to the {\it HEAO-1} A1 sample.          

%\com{Intro $B$G$O!"J}K!O@$K$D$$$F$N2r@b$OI,MW$J$$$G$7$g$&!#(B}

%The hard X-ray luminosity function of brightest AGNs are given by
%the HEAO-1 flux count rate and the intrinsic $N_{\rm H}$ obtained by 
%the detailed spectral results of ASCA and XMM-Newton.
%HEAO-1 flux was estimated by HEAO-1 observed count rate and the reduced 
%conversion factor of each source which included each spectral index and
%intrinsic $N_{\rm H}$ of the result of ASCA or XMM-Newton observation.
%HEAO-1 A1A3 (Grossan) count rate near above the source detection limit
%is assumed to be somewhat larger estimation than the true value, 
%we attempt to simulate the detected source counts
%and estimate the confusion effect. Then we re-calcurate the count rate of 
%each source, observed survey area of HEAO-1 A1A3 detector is build
%as a function of the confusion corrected flux.
%the luminosity function is calculated 
%from $\Sigma$ V$_{\rm a}^{-1}$ estimater. We provide that of X-ray absorbed
%AGNs which has the absorption column density 
%$N_{\rm H} > 10^{21.5} $cm$^{-2}$,
%and compare to that of X-ray unabsorbed AGNs.
%The spectral investigation of this complete sample will be one of the key
%constraints on answering the fraction of X-ray absorbed AGNs, and 
%verifying a tendency that there is no or only little absorption in 
%intrinsically high luminosity AGNs (Log Lx $> 44.5$ [erg s$^{-1}$]).
%We can provide the distribution of the slope of the underlying continuum,
%and the effect of warm absorbers or reflection components
%to the picture mentioned above.

 Throughout this paper, we use cosmological parameters 
(H$_0$, $\Omega_m$, $\Omega_{\lambda}$)$=$
(70 h$_{70}$km s$^{-1}$ Mpc$^{-1}$, 0.3, 0.7). When the units 
are omitted, $L_{\rm x}$ is measured in
erg s$^{-1}$ and $N_{\rm H}$ is in cm$^{-2}$.

\section{SAMPLE SELECTION}
\label{sec:Sample}

\subsection{Sample~1}

 As a part of our sample (Sample~1), we have used emission-line AGNs 
from \citet{Piccinotti1982}, with updated identifications in the on-line 
catalog (A2PIC) provided by  the High Energy Astrophysics Science Archive 
Research Center (HEASARC)\footnote{http://heasarc.gsfc.nasa.gov/}. 
We only use AGNs brighter than 1.25 (R15c/s) in the first scan,
which defines their complete flux-limited source list. 
Only one of the 66 sources above this count rate remains unidentified.
The limit corresponds to $\sim 2.7 \times 10^{-11}$ erg cm$^{-2}$ s$^{-1}$
in the $2-10$~keV band for a power-law spectrum with photon index 1.65.
%R15 count is a count rate derived using the 1.5$^{\circ}$ x 3.0$^{\circ}$
%FWHM fields of view of the second layer of the argon counter and
%both layers of a xenon counter in {\it HEAO-1} A2 experiment. 
In the \citet{Piccinotti1982}'s catalog, the regions 
between -20$^{\circ}$ and +20$^{\circ}$ in Galactic
latitude and the Large Magellanic Cloud (LMC) have been excluded to
minimize contaminations from Galactic sources. Also two AGNs which 
are within the region of the sky covered by Sample~2 (see below) have 
been excluded from Sample~1 (NGC~4151, NGC~5548). 
As a result of this exclusion, our Sample~1 consists 
of 28 AGNs with a survey area of $2.2\times 10^{4}$ deg$^2$.
 \citet{Malizia2002} pointed out that the $z=0.167$ AGN which was
listed in the on-line A2PIC catalog was a misidentification
of the X-ray source H~0917-074. Their {\it Beppo-SAX}/NFI observation 
found that the Seyfert 2 galaxy MCG~-1-24-12 at $z=0.0198$, which lies 
within the original {\it HEAO-1}/A2 error box of H~0917-074, dominated 
the X-ray emission in the FOV and had a much larger flux than 
the $z=0.167$ AGN. Therefore they concluded that MCG~-1-24-12 was the 
true counterpart of the {\it HEAO-1} A2 source. Thus we take MCG -1-24-12 
as the identification of H~0917-074.

\subsection{Sample~2}

 In order to increase the number of objects in the sample,
we have also utilized a deeper catalog of X-ray 
sources detected with the {\it HEAO-1} A1 experiment
by \citet{Wood1984}. Identifications of these sources
with the help of {\it HEAO-1} A3 experiment (modulation collimator) 
have been integrated as the ``MC-LASS catalog'' by R. Remillard. 
Both catalogs are available on-line from HEASARC (named A1 and A3 
respectively). Based on these 
catalogs, \citet{Grossan1992} defined a flux-limited subsample of 96 
AGNs (the Large Area Sky Survey/ Modulation Collimator identified 
sample of AGN or the LMA sample), in which he selected those with 
$>$0.0036 LASS cts s$^{-1}$ cm$^{-2}$, where the unit is defined 
in \citet{Wood1984}. 
\citet{Grossan1992} defined this limit to make a complete sample 
in $|b|>20\degr$. 

 Since it is not practical to make intermediate-resolution 
X-ray spectroscopic followup for all of the 96 AGNs, we have decided 
to limit the region for constructing a complete flux-limited sample. 
For this purpose, we selected a 55 degree radius region from the 
North Ecliptic Pole. This region was chosen to include the maximum 
number of AGNs which had already been observed or planned
to be observed by {\it ASCA} or {\it XMM-Newton}, i.e., 
to minimize the number of new observations, at the time of 
our initial {\it ASCA} AO-7 proposal. The defined area covers 
25 \% of the $\mid b \mid > 20\degr$ sky (5.1$\times$10$^3$ deg$^2$). 
Among the LMA AGNs within this region, we have excluded 3C 351 from our 
sample because it has been found to be confused with a BL-LAC object 
with a comparable flux from our {\it ASCA} observation.  
As a result, Sample~2 contains 21 AGNs brighter than the {\it HEAO-1} A1 
count rate greater than 0.0036 LASS c/s, which correspond to 
$\sim 1.9 \times 10^{-11}$ erg cm$^{-2}$ s$^{-1}$ in the $2-10$~keV band.
The overall identification completeness at this limit in our 
defined area is 85\%. However,  \citet{Grossan1992} argues, based 
on the comparison with {\it Einstein} and {\it ROSAT} data, that
most of the unidentified sources are Galactic stars (active coronae) 
or BL-Lac objects.
  
 Due to large beams and low count rate threshold, the {\it HEAO-1}
A1 count rate is subject to a bias due to confusion noise. 
We have made a first-order correction to this effect, as described 
in Appendix \ref{sec:Conf}.

\subsection{X-ray Spectroscopic Data and Sample Summary}

%All the \tm{AGNs in these} samples have to be observed with 
%ASCA or XMM-Newton to investigate the spectral analysis , which is not 
%reasonable with any expriments of HEAO-1 satellite to our study.

The observational data summary of each sources are listed in 
Tables~\ref{table:1-1} \& \ref{table:1-2}.
All of the AGNs in samples~1 and 2 are classified optically as 
emission-line Seyfert galaxies or quasi stellar objects (QSOs). Those
classified as BL Lac objects are excluded from our sample.

In order to obtain good quality spectra of all the AGNs in 
our samples, we observed the AGN which have not been previously observed 
nor scheduled with {\it ASCA}, {\it XMM-Newton}, nor {\it Beppo-SAX}.
We have made our own observations of five AGNs (Kaz~102, 3C~351, MKN~885,
H~1318+692, H~1320+551) with {\it ASCA} (AO7) 
and 2 AGNs (MKN~464 and H~0917+074) with {\it XMM-Newton} (AO1) to 
almost complete the spectroscopic study of the sample.
%We excluded 3C351 from our complete sample, which was found to be
%contaminated by a BL-Lac object with a comparable X-ray flux. 
Our  {\it ASCA} observation of 3C~351 and {\it XMM-Newton} data 
on H~0917+074 have not been used because of the reasons shown above.
We have used the spectral analysis result of MCG~-1-24-12 with 
the {\it Beppo-SAX} observation by \citet{Malizia2002} instead
of our {\it XMM-Newton} observation of H~0917+074.

\begin{figure}
%\plotone{plot-lb.eps}
\plotone{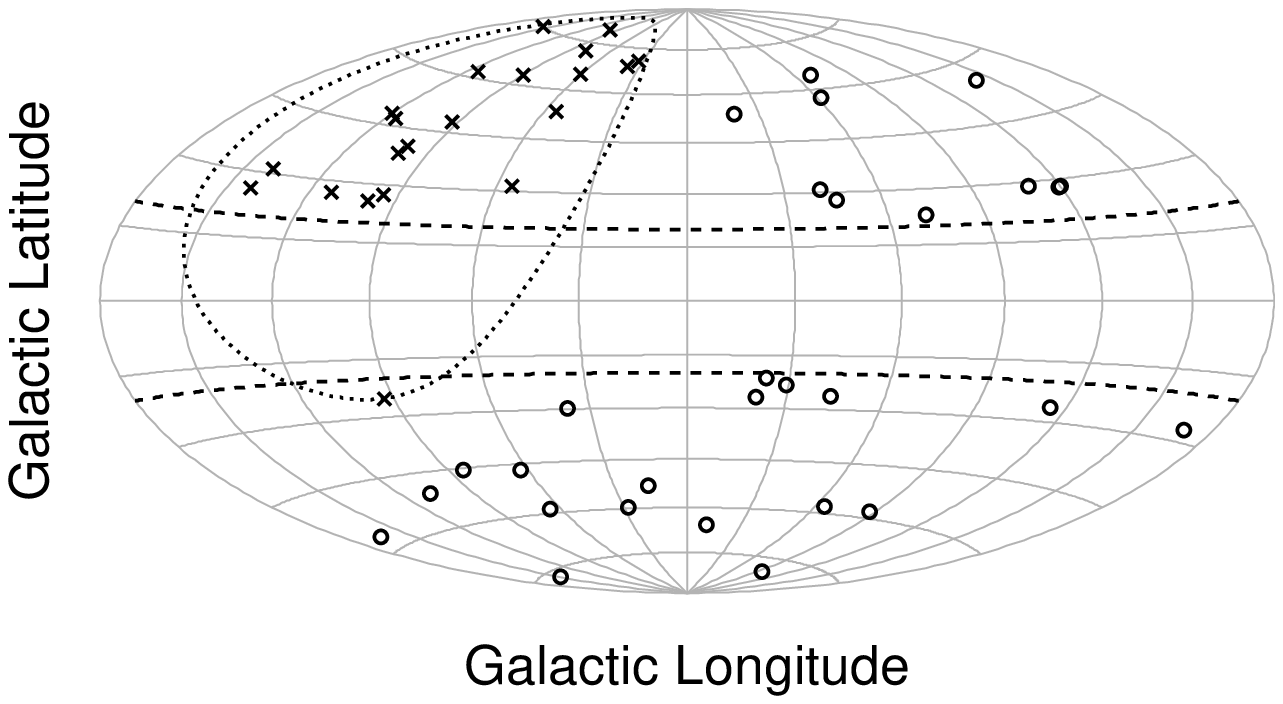}
\caption{Spatial distribution of our sample.
Dashed line represents b=$\pm$20 degree,
dotted line represents the selected region of Sample~2 (Grossan). 
Open circle~:~ Sample~1(Piccinotti),
Cross~:~ Sample~2(Grossan).
\label{SpatialDistribution}}
\end{figure}

\begin{figure}
%\plotone{hxlf_pigro2-6.eps}
\plotone{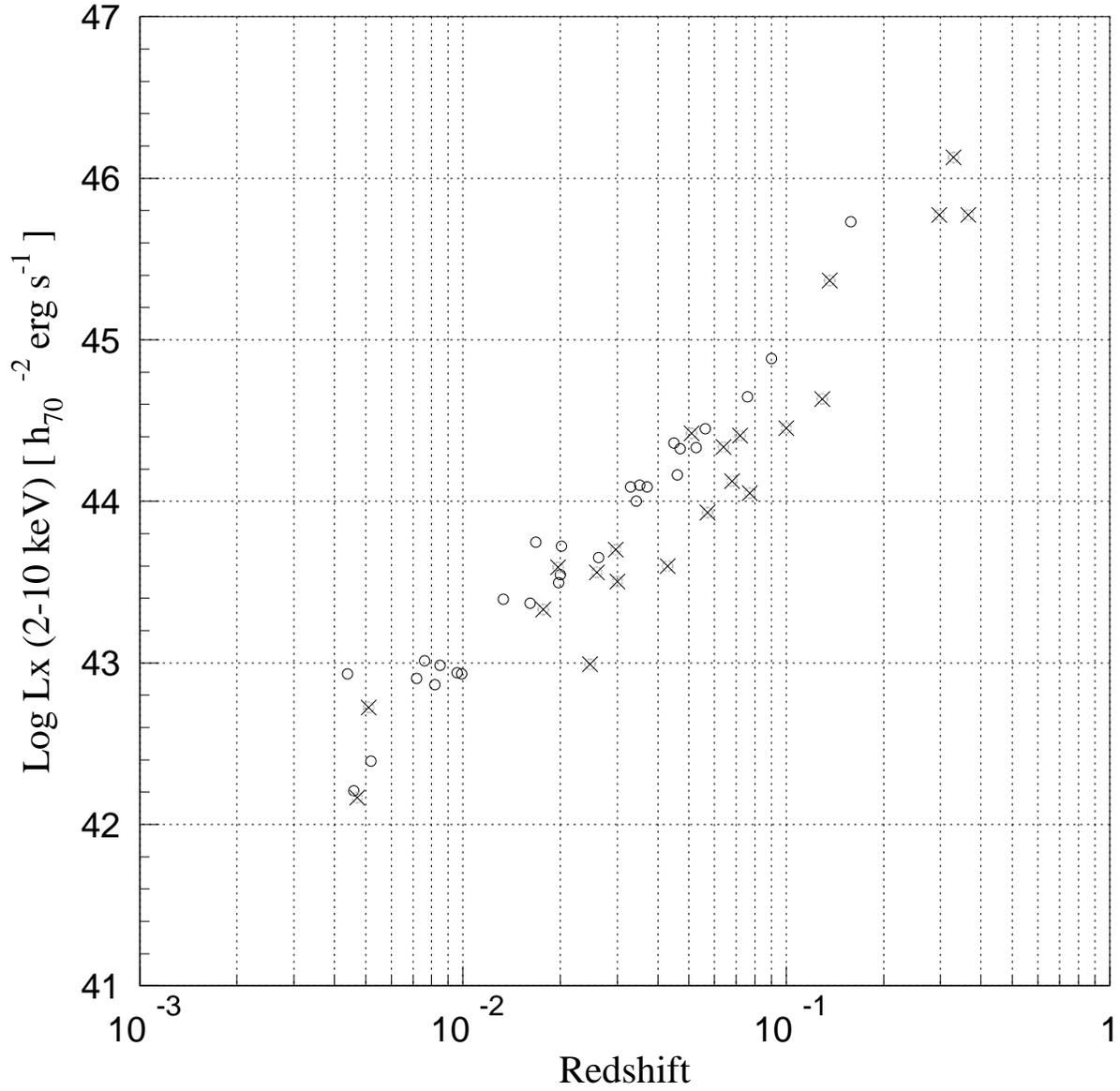}
\caption{Relation between X-ray luminosity and redshift of the sample.
Open circle~:~Sample~1 (Piccinotti). Cross~:~Sample~2 (Grossan).
For Sample~2, the flux is calculated from the corrected count rate described in 
Appendix~\ref{sec:Conf}.
\label{RedshiftAndLuminosity}}
\end{figure}

Figures~\ref{SpatialDistribution} and~\ref{RedshiftAndLuminosity} show
the spatial distribution and the relation between X-ray observed luminosity
and redshift of all the samples.
Putting both samples together, we have analyzed 21 sources of {\it XMM-Newton}
observation data (13 sources of Sample~1, 8 sources of Sample~2) and 14
sources of {\it ASCA} observation data (3 sources of Sample~1,
11 sources of Sample~2). For the other sources, we have
used spectral analysis results in literature. Among them, 
5 sources were observed with {\it XMM-Newton}, 6 with {\it ASCA}, 
and 2 with {\it Beppo-SAX}. 
Thus we get the spectral information on all but one (H~1537+339) AGN 
in the sample of 49 AGNs defined above. There is no {\it ASCA}, 
{\it XMM-Newton} or {\it Beppo-SAX} observation data for H~1537+339.
\citet{Remillard1993} showed an optical spectrum of the AGN identified 
with this X-ray source. It had broad-lines and a UV excess, which were 
typical of a type~1 AGN, and showed no sign of broad absorption 
lines (BALs). Thus we have assumed that it has an unabsorbed 
power-law spectrum ($\Gamma = 1.8$), which is typical of a type-1
AGN/QSO.

\section{DATA REDUCTION AND SPECTRAL ANALYSIS}
\label{sec:Analysis}

\subsection{Overall Procedure}

%\blank{$B$3$N(BParagraph$B$N5-=R$OI,MW$+!)(B}
 
% We basically use the spectral analysis results from literature 
%for the sources which have the complicated spectral feature. Those are
%NGC~7582, ESO~198-G24, 3C120, NGC~3783, NGC~3227, ESO~141-G55 and 3A~0557
%(\citealt{Turner1997,Guainazzi2003,George1998,Blustin2002,Gondoin2003a,
%Gondoin2003b,Turner1996}, respectively).
%MCG-6-30-15 has been extensively studied about a skewed 
%iron K$_{\alpha}$ line, reflection hump, and sharp drops in flux observed 
%in soft X-ray spectra were interpreted as absorption edges due to 
%O$_{\rm VII}$ and O$_{\rm VIII}$ in ionized gas within the AGN
%(\citealt{Reynolds1997,George1998,Lee2001,Turner2004}).
%NGC2992 and NGC7172 have a significant flux variability
%(\citealt{Reynolds1997,Turner1997}, respectively),
%NGC~526a is found to have apparent reflection
%components in hard X-ray band, and strong soft excess \citep{Turner1997}.
%NGC~4151 has long been known that it appears 
%to have a flat power law ($\Gamma$ $\sim$ 1.3 - 1.7), a substantial column 
%density of intrinsic absorbing gas, the complex iron K $\alpha$ emission line,
%and a strong soft excess at energies below 2 keV
%\citep{Schurch2002, Schurch2003}.

% Since we are not interested in such detailed structures,
%we take typical spectral indices and intrinsic
%absorption column densities from the literature for most of our
%AGNs.

We have reduced the {\it XMM-Newton} data using 
the Scientific Analysis System (SAS) ver 5.4.1. or later. 
We have compared the results of the reduction/analysis 
using SAS ver 6.0 and ``Current Calibration Files'' (CCF) updated 
as of December 2004 with those using SAS 5.4.1. for a few 
selected sources. The change of the results 
was insignificant for our purposes.  
About a half of the sources with available {\it XMM-Newton} data have
such high count rates that CCD pile up affects the spectral
analysis. In these cases, we extracted the source spectrum from
an annular region with inner and outer radii of about 12$\arcsec$ 
and 60$\arcsec$ respectively (exact radii vary)
to avoid the central peak. 
A background spectrum was taken from an annular region around the source 
with the inner radius of 1$\arcmin$, when the observation was made
with the  full window or large window mode. Small window mode
observations are usually made for bright sources and 
the background are negligible for these observations.  
%There were 5 sources which obviously had a time variability in the 
%light curve (NGC~7469, NGC~7314, ESO~103-G35, MKN~478, NGC~5548). 
%we address a whole phases of light curve to have a typical spectral shape
%about this sources. 
Since the {\it XMM-Newton} light curve of
NGC~5548 had a flare in the last 1/4 of the observation, we did not include 
this time interval into spectral analysis. In the spectral fitting, 
we used pulse-height channels corresponding to $0.4-10$~keV and $0.3-10$~keV 
for MOS and PN detectors respectively.

%NGC7469, NGC7314, ESO103-G35, $BIaDL$KJQF0!"A4BN$$$l$F$k(B
%IIIZw2, $BESCf$K(Bexcess$B$,$"$k!#$3$3$O$L$$$F$k(B
%MKN279, H1419+480, MKN6, $BESCf%]%m%]%m$H(Bexess$B$,$"$k!#$L$$$F$k(B
%MKN478, $BJQF0$b$"$k$,:G8e$@$1$-$?$J$$!#:G8e$NItJ,$L$$$F$k(B
%NGC5548, $B:G8e$K%U%l%"$_$?$$$JJQF0$,$"$k!#$3$3$O$L$$$F$k!#(B

We have used FTOOLS 5.0 and asca-ANL ver 2.00 to make data reduction and 
extraction of events for {\it ASCA} observation data.
The SIS and GIS pulse-height spectra have been created from events screened 
with the standard screening criteria. Source spectra has been extracted from
a 3$\arcmin$ radius circle
and background spectra has been accumulated from an off-source area of 
the same observation for SIS spectra. 
For the GIS spectra, we have used a 7$\arcmin$ radius region as a source
and the backgrounds from 8$\arcmin$ and 11$\arcmin$ annulus region centered 
on each source of the same observation.
The {\it ASCA} light curve of 
MKN~1152 had a flare in the middle and 
we have excluded the data during the flare. In the spectral fits, 
we have used pulse-height channels corresponding to $0.7-10$~keV 
and $1.0-10$~keV for SIS and GIS detectors respectively.

 For both {\it XMM-Newton} and {\it ASCA} data,  we have used XSPEC ver 11.0
or later for the spectral analysis.
%We need to derive the intrinsic absorption of each 
%AGNs to lead the direct power of the primary X-ray components
As our first step, we have fitted the reduced pulse-height
spectra with a single power-law and intrinsic absorption model 
(XSPEC model ``zwabs''). Galactic absorption (``wabs'') with
a fixed column density \citep{Dickey1990} has been always included. 
This first step analysis have been made in 
the $2.0-10$~keV range to avoid the effects of soft excess. We have also 
excluded the $5.5-8.0$~keV  range when there was a significant residual caused 
by absorption edge or 6.4 keV Fe K$\alpha$  line in the spectrum.
 The second step has been to add additional components, including 
a soft excess,  
an absorption edge, a reflection component, or a 6.4 keV Fe K$\alpha$ line 
to the spectral model and fitted to the full energy range whenever needed
to verify that the deviation from the simple best-fit absorbed power-law
does not have major effects on the spectral statistic and luminosity function 
analysis which are made later in this paper. We have also tried to 
include partial covering and/or ionized absorbers for some sources
as described in Sect. \ref{sec:SpecialAGN} .

%For example IC4329A, it is suggested to have a warm absorber and neutral 
%Fe K$_\alpha$ line. NGC4151 has a complex line feature, substantial column 
%density and a strong soft excess at energies below 2 keV.
%Many AGNs in our sample have soft excess features. Thus we cannot 
%explain most of them as a simple power-law component and a single
%absorption component. One AGN (Kaz 102) apparently showed an unusually
%hard single power-law spectrum with $\Gamma \sim 1$. This may be
%explained by a canonical AGN power-law $\Gamma\sim 1.4$ with an
%extreme warm absorber and/or reflection \citep{Miyaji2003}.  
%The detailed analysis of the spectral fitting on each sources are 
%shown in next paper.
% In the rest of our paper, however, we focus on the spectral
%index of the single power-law ($\Gamma$) and intrinsic absorption
%($N_{\rm H}$).

 The basic results of the simple absorbed power-law fit parameters
(photon index $\Gamma$ and intrinsic neutral absorber's column 
density $N_{\rm H}$) are summarized in Tables~\ref{table:2-1} 
\& ~\ref{table:2-2}. These numbers are from our first-step 
analysis or similar simple absorbed power-law fits from 
literature except for those detailed in Sect. \ref{sec:SpecialAGN}.

\citet{Piccinotti1982}, used the conversion factor of one R15 
count s$^{-1} \sim 2.17 \times 10^{-11}$ ~erg cm$^{-2}$ s$^{-1}$ 
in the $2-10$~keV energy band for a power-law spectrum with photon 
index 1.65, to derive the flux for most of their AGNs. Since we 
have spectral information for each object, we have re-calculated 
the fluxes of Sample~1 objects from the R15 count rate using the 
actual {\it HEAO-1} A2 response function and our spectral results. 
The corrected observed flux is listed as the {\it HEAO-1} flux
in Tables~\ref{table:2-1}\& \ref{table:2-2}.
An underlying assumption is that $\Gamma$, intrinsic $N_{\rm H}$ and 
other spectral parameters (except the global
normalization) of an AGN do not change with the variation of its luminosity.
For Sample~2, \citet{Grossan1992} used conversion factor 
$5.29 \times 10^{-9}$ ~erg cm$^{-2}$ s$^{-1}$/LASS count s$^{-1}$
in the $2-10$~keV energy band appropriate for a power-law spectrum 
with photon index 1.7 for each flux.
We have re-calculated the fluxes using our spectroscopy results 
for Sample~2 as well.

 The luminosities listed in Tables~\ref{table:2-1} 
\& \ref{table:2-2} are  intrinsic luminosities, i.e., 
absorption-corrected luminosity using the fitted N$_{\rm H}$. 
We show the absorption-corrected  luminosities because we are 
more interested in the intrinsic AGN power in the later analysis.

%$B$3$NItJ,$OITMW!#:G8e$N(Bdiscussion$B$N$H$3$m$K(B caveat $B$H$7$F=q$$$F$*$$$F$b(B
%$B$$$$$,!"!"!"(B
%
%The sources which has less than 0.015 of redshift should be excluded,
%because it may affect uncertainty or have other (gravitational ?) 
%potencial effects. However, we research to provide the bright-end 
%constraint of HXLF and the AGN population synchesis model about cosmic
%X-ray background with the statistically analysis with a 
%significant number of sources, so that we added such sources.

\subsection{AGNs with Special Considerations}
\label{sec:SpecialAGN}

 In this subsection, we summarize the spectral features
of the AGNs for which the deviation from the simple power-law
absorbed by neutral gas has a major impact on discussing the 
spectral statistics, intrinsic luminosities, and absorbing 
column densities. In particular, we have taken
special attention to AGNs which gave unusually hard power-law
index ($\Gamma \la 1.4$) in the first step simple absorbed power-law 
fits, since they are likely to be heavily affected by partially covered 
and/or warm absorbers. Six AGNs in the entire sample 
have been found to fall in this category as detailed below.  

 In calculating the intrinsic luminosity,  we have 
used the spectral models involving  partial and/or ionized absorbers. 
The global normalization has been adjusted to give the 
observed {\it HEAO-1} A2 (A1) count rate, if the AGN belongs to 
sample~1 or sample~2. The intrinsic luminosity in Tables ~\ref{table:2-1} 
\& \ref{table:2-2} have been corrected for all absorbers.
  For the column density in these 
tables, we have used the sum of all neutral absorbers (including 
partial absorbers) as a representative value, but have not included those 
of any warm (ionized) absorber components. We explain our treatments for
these 6 AGNs below.

\begin{description}

\item[NGC~7582] A simple absorbed power-law fit to the {\it ASCA} data
  gave $\Gamma = 1.2$ and $N_{\rm H}=6\times 10^{22}$ ~cm$^{-2}$. The best
  description of its X-ray spectrum was given by \citet{Turner2000} from
  the {\it Beppo-SAX} data. Their model included an intrinsic power-law
  component of $\Gamma = 1.95$ absorbed by two components:
  a totally covering $N_{\rm H,22} = 14.4$ absorber and a 
  thicker absorber with $N_{\rm H,22} = 160$ blocking $\sim 60\%$ of the
  X-ray source.  We employ this model for de-absorption (see above) and
  the sum of these two absorbing columns for the representative neutral
  column density. 

\item[3A~0557-383] The hard X-ray source 3A~0557-383 was identified with a 
 Seyfert 1 galaxy. \citet{Turner1996} reported the complex absorption 
 below 2~keV of the X-ray spectrum from an {\it ASCA} observation. They 
 found a good fit with either 96 \% of the source covered by a column of 
 3.1 $\times$ 10$^{22}$ ~cm$^{-2}$ of low-ionization gas or full covering 
 by a column of 3.6 $\times$ 10$^{22}$ ~cm$^{-2}$ of highly ionized gas, 
 and an neutral iron K-shell emission line with equivalent width ~300 eV.
 We used the latter model (Model D of \citealt{Turner1996}).

\item[Kaz~102] The results of detailed analysis of our own {\it ASCA} 
 data of this object have been reported by \citet{Miyaji2003} (M03). 
 Kaz~102 has no sign of heavy absorption or a Fe K$\alpha$ line. A single 
 power-law with $\Gamma \sim 1$ described the {\it ASCA} spectrum well,
 while it had softer spectra when it was observed by {\it ROSAT}.  
 This hard spectrum is probably due to extreme warm absorber
 component and/or reflection as discussed in M03. In this
 study, we employ the warm absorber model with intrinsic 
 $\Gamma\sim 1.9$ in M03 for absorption correction. No 
 neutral absorber is detected in the analysis and thus we assumed 
 a zero $N_{\rm H}$ for the neutral gas.

\item[3C~445] The nearby broad-line radio galaxy (BLRG) 3C~445 
  has very hard power law $\Gamma$=1.08 and a large intrinsic absorption
 ($N_{\rm H} \sim 1.4 \times 10^{23}$ cm$^{-2}$) in our first-step analysis
 of the {\it XMM-Newton} observation data. In \citet{Sambruna1998}, 
 the X-ray spectrum of 3C~445 observed with {\it ASCA} were described 
 by a dual absorber with an intrinsic flat power-law ($\Gamma \sim 1.3$) 
 continuum. 
 We have made our own analysis of the {\it XMM-Newton} data with two 
 intrinsic absorbers, neutral partial covering and full covering cold gas 
 in $0.6-12$~keV band like Model C of \citet{Sambruna1998}. Our analysis show  
 a hard intrinsic power law ($\Gamma$ =1.11$\pm$0.05) absorbed by  
 partial covering ($90\pm 10$\%)
 neutral gas with $N_{\rm H} \sim 1.32 \pm 0.06\times 10^{23}$ 
 cm$^{-2}$ and no full covering cold gas
 ($\chi^2/d.o.f = 1.27/159$).

 We employ our {\it XMM-Newton} result for intrinsic luminosity
 calculation and statistics. Taking \citet{Sambruna1998}'s model had 
 negligible effects on the analysis in next sections. 

\item[MKN~6] 
%% Our first step ($2.0-10$~keV) analysis of the {\it XMM-Newton} 
%% spectrum gave a hard power-law  ($\Gamma$ = 1.33) absorbed by an 
%% $N_{\rm H} \sim 3.2 \times 10^{22}$ ~cm$^{-2}$ neutral absorber. 
 A detailed spectroscopic analysis of this object
 is given by \citet{Stefan2003}, who analyzed the same {\it XMM-Newton} data 
 combined with {\it Beppo-SAX} data. Their joint analysis 
 preferred a model (Model 4) with a fixed-index power law ($\Gamma$=1.81) 
 and two partially covering neutral absorbers 
 (56$\pm$2~\% of 8.1$\pm 0.5 \times 10^{22}$ ~cm$^{-2}$ and
 91$\pm$1~\% of 2.3$\pm 0.1 \times 10^{22}$ ~cm$^{-2}$), with a very
 small reflection component. We have re-analyzed the {\it XMM-Newton} 
 using this model and checked the spectral fitting results, when $\Gamma$ 
 was set to free parameter. We found a slightly harder spectral index and 
 small column density, but the difference of derived intrinsic X-ray 
 luminosity is less than 10\%. This has negligible effects on our 
 HXLF and other statistical conclusions. We listed the sum 
 of column densities of the dual absorbing gas and $\Gamma =$ 1.81 and used 
 the spectral shape of  \citet{Stefan2003}'s Model 4 for the absorption
 correction.

\item[NGC~4151] This Seyfert 1.5 galaxy NGC~4151 has long been known that 
 it appears to have a flat power law ($\Gamma$ $\sim 1.3 - 1.7$), a 
 substantial column density of absorbing gas (N$_{\rm H}$ $\sim$ 
 10$^{23}$ ~cm$^{-2}$), a strong soft excess at energies below 2 keV,
 and the broad iron K $\alpha$ emission line.
 We take the result of Model 2 in \citet{Schurch2003}, who modeled 
 the spectrum with two absorption components, a partially 
 photoionized absorber and a neutral absorber, and a neutral reflection 
 continuum. The column density of the neutral absorber is listed in 
 Table~\ref{table:2-1} and the absorption correction has been made
 with their Model 2.
\end{description}

\subsection{Other Spectral Features}

\begin{figure}
\includegraphics[angle=270,scale=0.60]{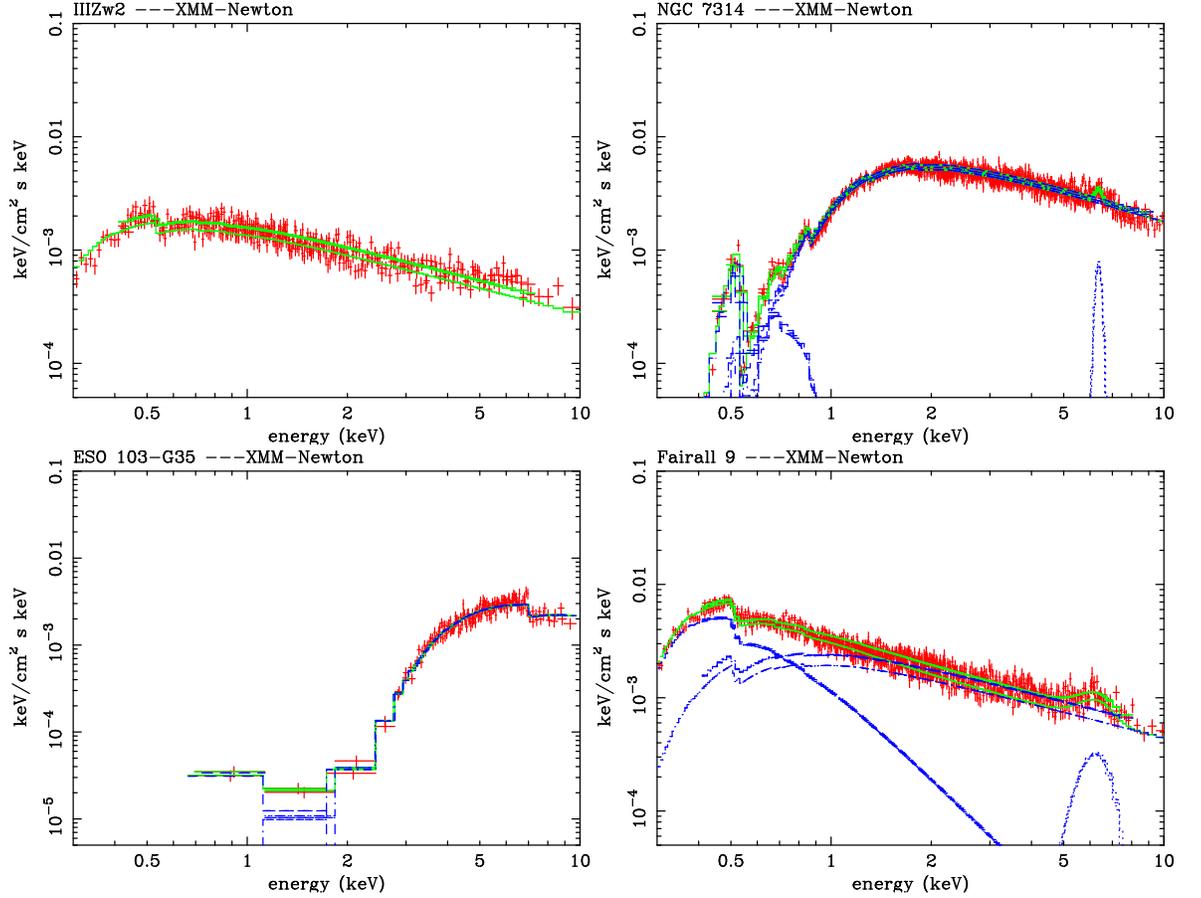}
%\plotone[angle=90,scale=0.50]{eufplot6.eps}
%\plotone{f3.eps}
\caption{Examples of spectral fit results in the $E\;I(E)$ scale, 
for IIIZw2, NGC~7314, ESO~103-G35, and Fairall~9 in 0.3-10~keV band.
The spectra are represented by thin red crosses, and green and blue lines
represent the all model spctrum and each spectral component, respectively.
The details of each spectral component are listed 
in Table~\ref{table:2-1}-\ref{table:4}.
Figures 3.1 - 3.5 are available in the electronic edition of the Journal.
\label{Spectralfit}}
\end{figure}

We have also examined the spectral fits in the broad band 
spectrum more in detail, in particular, paying attention to the 
existence of iron lines and/or a soft excess. We have started 
with a power-law with the photon index fixed to that in first step 
analysis, with the Galactic absorption at the AGN position. For 
those where the simple absorbed power-law model have significant residuals
in the broad-band spectrum, we considered more complicated models,
involving, e.g., a soft excess and or an iron line.
The fitting results with the soft excess are shown in Table~\ref{table:3}. 
We listed the only best fit results for soft excess, within 
several spectral model fittings (blackbody, 
broken-power-law, 2-powerlaw, soft thermal emission, dual absorber, etc.. ).
Typical spectrum of the sample in our spectral fitting are shown in 
Figure~\ref{Spectralfit}. 
For those the addition of an iron line improved the fit, 
the central energy, normalization, and equivalent width the 
line, represented by a single Gaussian, are listed in 
Table~\ref{table:4}. In the spectrum of MKN464, there are 
significant residuals in $1.5-3.0$~keV band and this range has been 
excluded for the analysis.

The photon index and intrinsic column 
density for most of all AGNs are consistent to that in first step analysis
within errors, even if the photon index and the intrinsic absorption 
($N_{\rm H}$ in ``zwabs'') are set to free parameters in this 
second-step spectral analysis. Thus the existence of soft excess and iron 
emission line does not have significant effect in our 
further analysis. 

The statistics of such features has its own importance.
However, in our current data, the limits on the detection of these features 
vary observation by observation and difficult to model. Thus one should not 
take advantage of the good completeness of our sample to make a statements 
on the statistical properties of the iron lines and soft excess in combination
with our sample definition.

\section{STATISTICS OF SPECTRAL PROPERTIES}
\label{sec:Resultspectrum}

\subsection{Neutral $N_{\rm H}$ and Photon Index Distribution}

\begin{figure}
%\plotone{hxlf_pigro2-1.eps}
\plotone{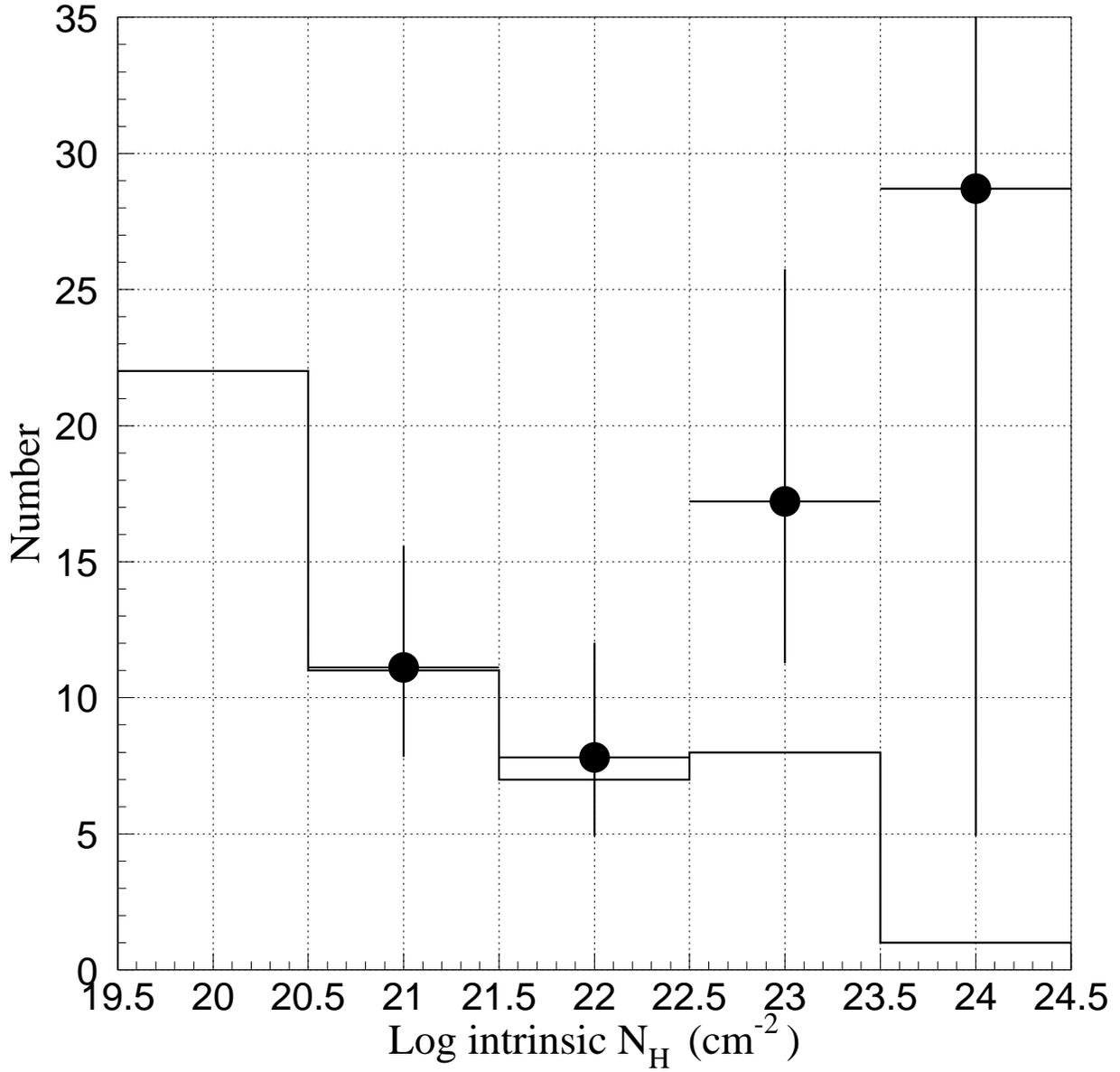}
\caption{The intrinsic N$_{\rm H}$ (neutral) distribution of the AGNs in our sample. 
 We also plotted spatial number density calculated from Eq.~\ref{SpatialDensity}.
\label{IntrinsicNH}}
\end{figure}

Figure~\ref{IntrinsicNH} shows the distribution of the absorbing
column densities (N$_{\rm H}$) of neutral absorbing gas intrinsic
to the AGNs. 
%intrinsic absorption which is lower than Log $N_{\rm H}$
%[cm$^{-2}$] $< 20.0$, because we must have the spectral analysis
%after subtracting the absorption of our galaxy, which has about
%Log $N_{\rm H} \sim 20.0$ in the most of the source's line
%of sight.
%So we took Log $N_{\rm H} = 20.0$ the lower limit of 
%the intrinsic column density for our study.
As described in Sect. \ref{sec:SpecialAGN},
we have added the value of any partial covering neutral column 
densities to the $N_{\rm H}$ value used here.
On the other hand, the column densities of ionized absorbing gas have 
not been included (3A~0557, Kaz~102, NGC~4151).
This is because the neutral gas and ionized gas come
from different regions, the former being in a molecular torus at about
$10^{18}-10^{20}$ cm from the nucleus, while the latter at
a region closer to the nucleus ($\sim 10^{17}$ cm, 
(e.g. \cite{Reynolds1995}). Thus they do not have much relevance
to each other.
%There is only one source, NGC7582 in $23.5 < \log\,N_{\rm H} < 24.5$.
%There are 12, 5, 9 sources in 
%$\log\,N_{\rm H} \sim 20.5-21.5$, $21.5-22.5$, $22.5-23.5$ ~cm$^{-2}$ bin
%respectively.

We also plot an expected relative spatial number density in each 
$N_{\rm H}$ bin, assuming a Eucleadian $\log\,N  - \log\,S$ 
relation for the intrinsic flux distribution of the X-ray AGNs: 
\begin{eqnarray}
N_{\rm spatial}=\left(\frac{f_{\rm intrinsic}}
{f_{\rm absorbed}}\right)^{1.5} \times N_{\rm raw}
\label{SpatialDensity}
\end{eqnarray}
in which, $N_{\rm spatial}$ is spatial number density, 
$f_{\rm intrinsic}$ is intrinsic flux of primary X-ray component,
$f_{\rm absorbed}$ is observed flux (after absorption), and
$N_{\rm raw}$ is the raw number in the $N_{\rm H}$ histogram.
Upon calculating the $f_{\rm intrinsic}$ of each bin,
we have assumed that the intrinsic power-law source of $\Gamma = 1.7$
was absorbed by cold gas of mean column density of each bin. 
 This may not be correct in case of the source has partially covering 
absorbers. However, the effect is much smaller than the errors 
due to small number statistics in each bin.

%\com{$B$3$N(BPara$B$O!"(BHXLF$B$N7W;;$N$H$-(B}

% We note that 3 sources changed the classification 
%from unasborbed to absorbed AGNs used without adopting the further 
%analysis discussed in Sect .~\ref{sec:SpecialAGN}.

\begin{figure}
%\plotone{gamma.eps}
\plotone{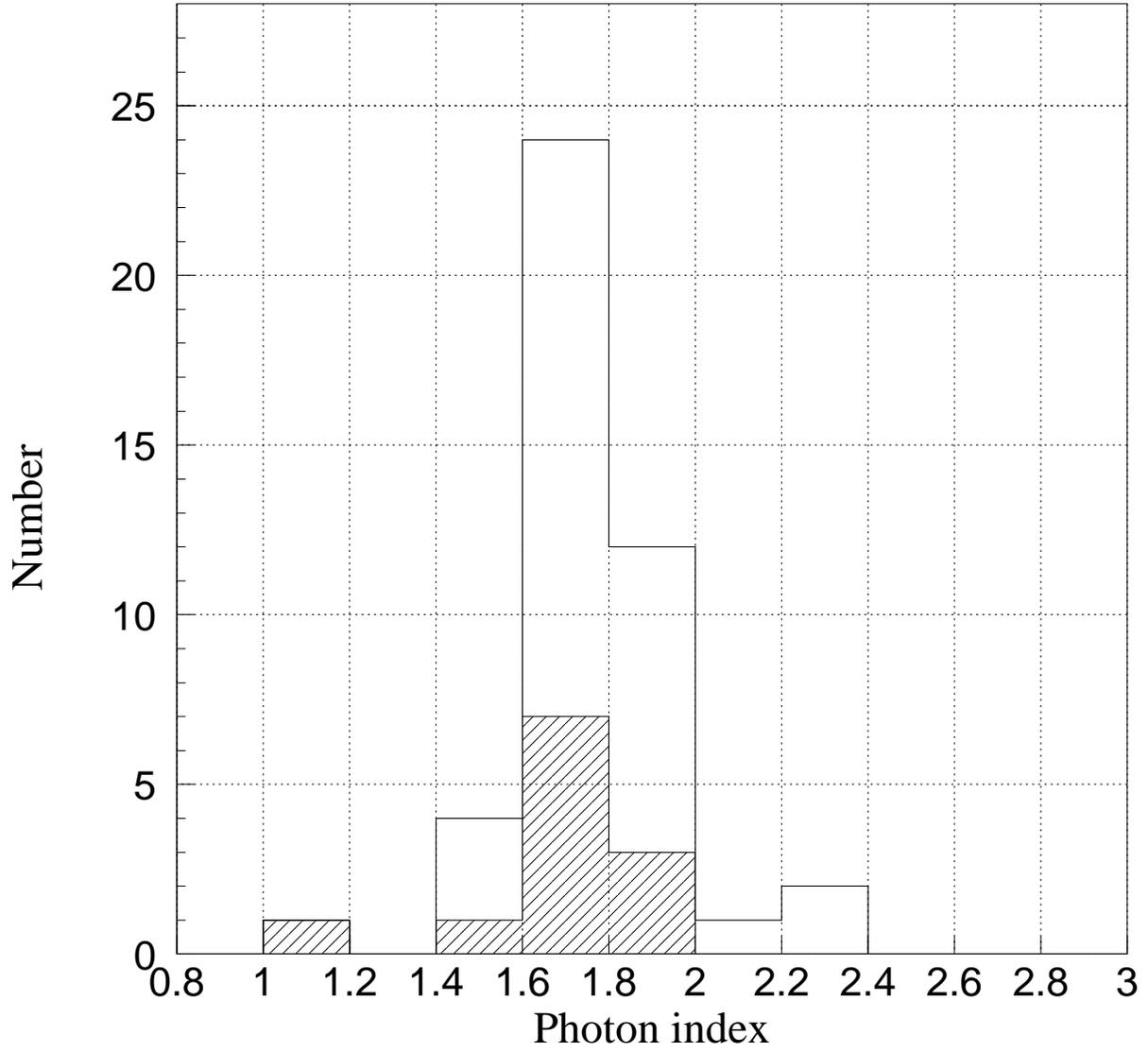}
\caption{Photon index distribution of the sample. 
The open histogram represents all the sample, the filled histogram represents
X-ray absorbed AGNs.
\label{PhotonIndex}}
\end{figure}

 Figure~\ref{PhotonIndex} shows the distribution of spectral photon index.
The $\Gamma$ values of NGC~3227, 3A~0557, Kaz~102, H~1537+339, and MKN~6 are 
excluded from the histogram, because they have been fixed during 
the spectral fit, or no detailed spectral information.
The mean photon index is 1.76 with $\sigma = 0.2$.
This is similar to the result of \citet{Williams1992}, who studied an 
incomplete sample of 13 QSOs observed by {\it GINGA} and 
obtained the result of the mean photon index of 1.81$\pm$0.31. 
%We point out the extreme location of 
%\citep{Miyaji2003} in this figure, if this source were throught to be
%no intrinsic absorption and extremely hard spectral index 
%($\Gamma =1.02\pm0.06$ in our first analysis).
Even if we use the results from first-step single absorbed power-law
fit results (Sect.~\ref{sec:Analysis}) the 6 sources
discussed in Sect.~\ref{sec:SpecialAGN}, the mean photon index 
has not change remarkably, giving 1.72$\pm $0.24.
In \citet{George2000}, the mean photon index has $\sim 2.0$
for various spectral analysis, while the relatively wider distribution 
($1.3 < \Gamma < 3.0$). And in \citet{Reeves2000}, they derived mean photon 
index $1.89\pm0.05$ with $\sigma = 0.27$. The two other observations are
somewhat softer distributions, which is assumed be due to the effect of 
a reflection component.
We have found no correlation between $\Gamma$ and 
hard X-ray luminosity, which is consistent with results of \citet{George2000}, 
and \citet{Reeves2000}.

\subsection{Relation between Hard X-ray Intrinsic Luminosity and $N_{\rm H}$}

\begin{figure}
%\plotone{hxlf_pigro2-3.eps}
\plotone{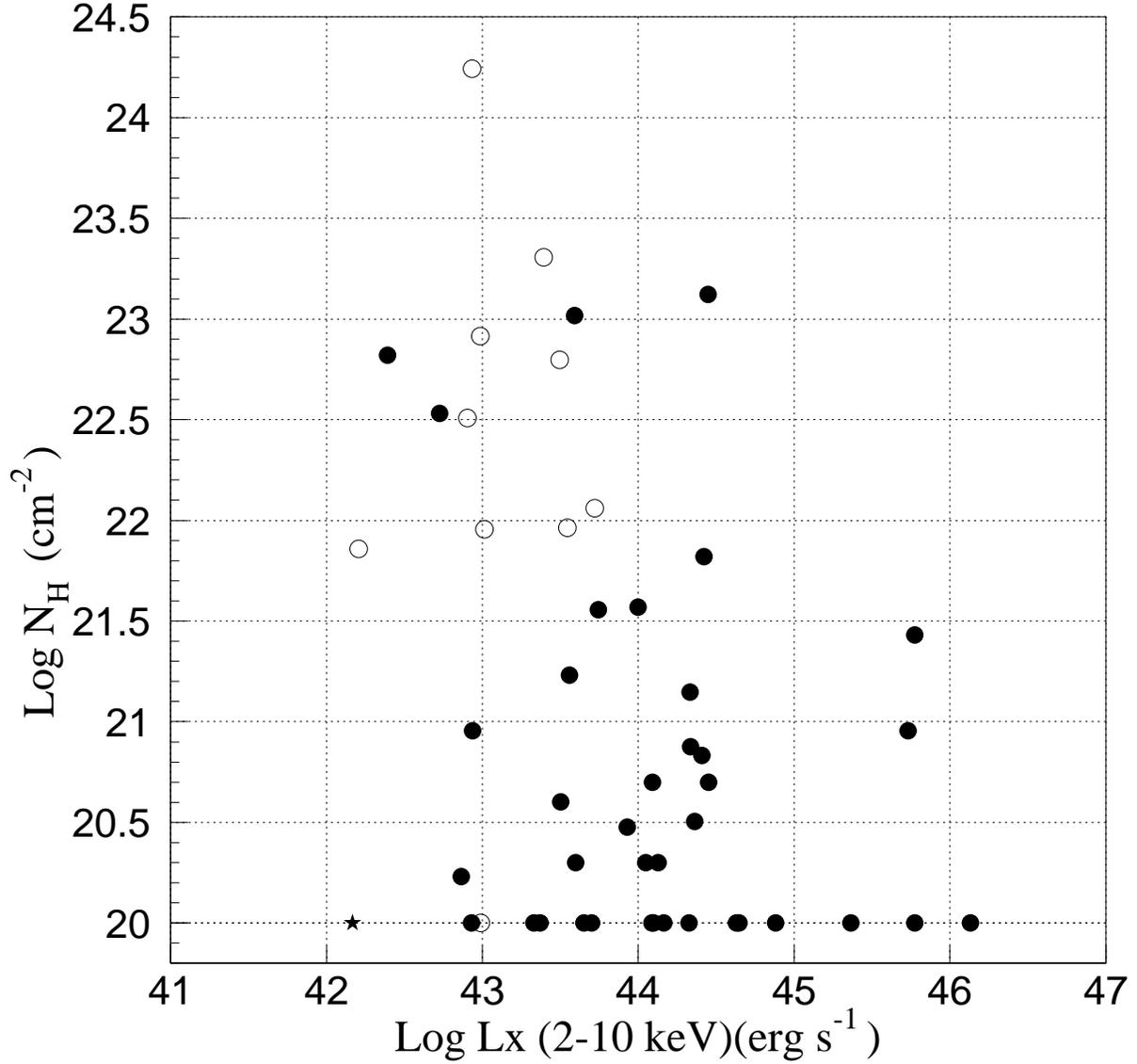}
\caption{Relation between hard X-ray luminosity and intrinsic N$_{\rm H}$.
Type~1 galaxy (Seyfert 1, 1.5, Type~1 QSOs)
are shown as filled circles and Type~2 AGNs (Seyfert 1.9 - 2)
are shown as open circles respectively.
One probable  LINER is plotted with a filled star.
The luminosity is intrinsic (before absorption).
\label{RelationBetweenLAndNH}}
\end{figure}

 Figure~\ref{RelationBetweenLAndNH} shows the 
scatter of the AGNs, in the hard X-ray luminosity -- 
intrinsic neutral $N_{\rm H}$ plane.
Different symbols are used for AGNs with different 
optical types, i.e. type 1 (Seyfert 1-1.5, type1 QSO, or BLRG),  
type 2 (Seyfert types above 1.8) AGNs or LINER candidate.
Most of optical type 1 AGNs have small X-ray absorption, 
$\log\,N_{\rm H} \le$ 21.5, and most of Seyfert 2 galaxies have large 
X-ray absorption, $\log\,N_{\rm H} \ge$ 21.5 as widely known.
 Some AGNs optically classified as type 1 have large $N_{\rm H}$ 
($\log\,N_{\rm H} \ga$ 21.5) values.  They usually have X-ray spectra 
involving partially covering absorbers. 
We have include such sources in 
the absorbed AGN group for the HXLF analysis.
The optical classifications in this paper are not necessarily  
accurate, because they are based on the detection significance of broad 
emission lines and depend on the quality of the optical spectra.
Indeed, there are some AGNs (ESO103-G35, NGC526a, 3c445, etc..)
for which different sources give different optical classifications.
However, our spectral results may suggest that the anomalous cases reflect the 
existence of substantial variance in the dust to gas ratio, or
a geometrical separation to the line of sight. 
More significant data analysis with careful regard to the variability
would derived whether these objects are common in AGNs or not.

%\chg{$B$$$i$J$1$l$P>C$7$^$9!#(B TM:$B$=$&$G$9$M!"$A$g$C$H>iD9$G$9!#>C$7$^$7$g$&!#(B
%In this paper, the results of 6 sources were 
%apparently different from
%that of previous study U03, then 5 sources were converted to the group of 
%absorbed AGNs while 1 sources was converted to that of unabsorbed AGNs. 
%This indicated that more than 10~\% of sources (about 30~\% of absorbed AGNs)
%were misjudged about the $N_{\rm H}$, and the detailed spectral analysis 
%was needed to divide the sources into even only two bins.}

 We observe no AGN with $\log\,L_{\rm x}>44.5$ and $\log\,N_{\rm H} \ga$ 21.5.
This is consistent with U03's analysis including {\it ASCA} and
{\it Chandra} surveys, which find a decrease of absorbed AGN 
fraction towards high intrinsic luminosities. A further discussion
of this effect will be given in Sect.~\ref{sec:Disc}. 

% As this result, we would be able to detect about 9-10 sources in this region
%Log L$_{\rm X} >$ 44.5 and Log N$_{H} >$ 21.5, even if we added the effect
%of large intrinsic absorption to the Soft X-ray luminosity function,
%with which we probably miss a source.
%it suggested that, absorbed AGN's luminosity function is apparently different
%from unabsorbed AGN's luminosity function.
%We can obviously indicate the implication about brightest AGNs with 
%this complete samples.

%%($B:#$^$G$b(Bluminosity$B$,Bg$-$$$H(BNH$B$NBg$-$J$b$N$,$$$J$$$H$$$&$3$H$O(B
%%$B$$$/$D$+<(:6$5$l$F$$$?O@J8$r8fB8CN$G$7$g$&$+(B?)
%%$B$3$3$OM=B,$5$l$k8D?t$N6qBNE*$JF3=PJ}K!$b=q$$$?$[$&$,$h$m$7$$$N$G$7$g$&$+$M!#(B

%\com{$B2<$O(BDiscussion$B$K(B}
% However, as well as U03, our classification using optical
% spectral data are heterogeneous, not classified accurately, and not have 
% an identification with all the sample because of the quality of the 
% optical spectra.

\section{LOCAL HARD X-RAY LUMINOSITY FUNCTIONS}
\label{sec:HXLF}

 Our purpose of this section is to calculate the local  
Hard ($2-10$~keV) X-ray Luminosity Function (HXLF) of AGNs
from our sample.
We have taken the U03's approach and constructed the $2-10$~keV 
HXLFs for the {\em intrinsic} X-ray luminosity. 
We have divided the sample into unabsorbed and absorbed AGNs 
at $\log\,N_{\rm H}= 21.5$ for the use with HXLF analysis. The number 
of X-ray absorbed AGNs is 16, while that of X-ray unabsorbed 
AGNs is 33 in our sample.

 We have used the $\Sigma V_{\rm a}^{-1}$ estimator \citep{Avni1980},
which is a modified version of original $\Sigma V_{\rm max}^{-1}$ 
\citep{Schmidt1968}. We can write the binned HXLF as,
\begin{eqnarray}
\frac{d\Phi}{d\log\,L_{\rm x}}
(\log\,L_{\rm x}) \approx
\Sigma_i \frac{1 / V^{i}_{\rm a}}{(\Delta \log\,L_{\rm x})}
\label{eq:sigmaVa}
\end{eqnarray}
where $V^{i}_{\rm a}$ is the available comoving volume 
in which the $i$-th AGN would be in the sample, within the redshift 
range. The $V^{i}_{\rm a}$ value can be expressed by
\begin{eqnarray}
V^{i}_{\rm a} = 
\int^{z_{\rm max}}_{z_{\rm min}} d_{\rm A}(z)^2 c\frac{d\tau}{dz}(z)
\cdot A_{\rm eff}({\rm CR}) ~{\rm d} z
\label{eq:Va}
\end{eqnarray}
where $d_{\rm A}(z)$ is the angular distance, $d\tau / dz (z)$ is the 
differential look back time per unit z, CR is the observed 
count rate of the AGN if it was placed at the redshift $z$, and 
$A_{\rm eff}({\rm CR})$ is the survey area as a function of the 
count rate. We take $z_{\rm min}$ = 0 and $z_{\rm max}$ = 0.4 to
include all the AGNs in the sample.

 Since our sample consists of two subsamples defined in the 
count rate limits from two different detectors, our $V_{\rm a}$
consists of two parts:  
\begin{eqnarray}
V^{i}_{\rm a} = V^{i}_{\rm a,1} + V^{i}_{\rm a,2},
\label{eq:Va1+Va2}
\end{eqnarray}
where $V^{i}_{\rm a,1}$ and $V^{i}_{\rm a,2}$
are the comoving volumes of the universe within which the $i$-th AGN 
would be observable in the Sample~1 and Sample~2 respectively.
In calculating $V^{i}_{\rm a,2}$ of an AGN in Sample~1,
we have used the predicted LASS ~cts s$^{-1}$ cm$^{-2}$ of {\it HEAO-1} A1 
experiment, which has been estimated by using the {\it HEAO-1} A2 R15CR, 
the spectral shape of this source, and the detector responses of {\it HEAO-1} 
A2 experiment, and vice versa. In the HXLF calculation, we have used 
the observed flux (i.e. after absorption) in estimating the $V^{i}_{\rm a}$, 
while X-ray intrinsic luminosity (de-absorbed luminosity) has been used to 
divide the sources into luminosity bins.

 As mentioned in Sect.~\ref{sec:Sample}, we have used the {\it effective} 
survey area for Sample~2 after correcting the LASS count rate 
for the effects of confusion noise. This correction is discussed in detail 
in Appendix~\ref{sec:Conf}. After the correction, the area as a function
of the {\em corrected} count rate gradually decreases as the $CR_{\rm corr}$
decreases to the limit of  1.9 $\times 10^{-3}$ LASS cts s$^{-1}$ cm$^{-2}$. 

 A small number of very luminous AGNs in our sample are observable
at redshifts where evolution has a non-negligible effect 
($z \gtrsim  0.2-0.3$). 
Thus we have also derived the {\em evolution-corrected} HXLF by giving
weights to the  $V^{i}_{\rm a}$ calculation, 
\begin{eqnarray}
V^{i \prime}_{\rm a} = 
\int^{z_{\rm max}}_{z_{\rm min}} d_{\rm A}(z)^2 c\frac{d\tau}{dz}(z)
\cdot A_{\rm eff}(CR) ( \frac{1 + z}{1+z_0} )^p ~{\rm d}z,
\label{eq:Va_evcorr}
\end{eqnarray}
with which we can estimated the HXLF at $z = z_0$  by
including the density evolution factor $[(1+z)/(1+z_0)]^p$. 
We use the index $p = 5$, which is consistent with the soft X-ray
luminosity function evolution by \citet{Hasinger2005}. The impact
of the evolution to the XLF was found to be as small as $\sim 10\%$ in the  
$45.5 < \log\,L_{\rm x} < 46.5$ bin. 
Since our sample is mainly local, K-correction has a negligible effect. 
Most luminous AGNs are unasborbed with the photon index $\sim$ 1.9,
with very little K-correction even at $z=0.4$. 
Absorbed AGNs typically have such low luminosities that they 
are not detectable at high enough redshifts for the K-correction
to be important. The K-corrections are typically $\lesssim$ 5\%
for the absorbed AGNs.

Figure~\ref{HXLF} shows the Hard X-ray luminosity function 
calculated from our sample for the evolution-corrected case.
%Figure~\ref{HXLF}~(a) shows the 
%raw HXLF, while the evolution-corrected HXLF is shown 
%in Figure~\ref{HXLF}~(b). 
We take $z_0=0$ for Figure~\ref{HXLF}.
We have made the luminosity bins same in all HXLF calculations
for comparisons.
We only have upper limits for absorbed AGNs in the three highest 
luminosity bins and we plot only interesting (90\%) upper limits, 
i.e. those which make significant constraints. 
Figure~\ref{HXLF} clearly shows different X-ray luminosity function slopes 
between absorbed and uabsorbed AGNs at the high luminosity end. 
The ratio of absorbed / unabsorbed XLF is 2.35$\pm$2.1 in 
$\log\,L_{\rm x} \sim 42 - 44$, and 0.11$\pm$0.08
in $44 - 44.5$). The significance of the difference of the
absorbed/unabsorbed ratio in these two luminosity bins has been
estimated using a bootstrap resampling. Out of all the 41 AGNs in
the $\log\,L_{\rm x} \sim 42 - 44.5$ range, we have randomly resampled
the same number of AGNs allowing duplicates. The ratio of the  
absorbed / unabsorbed HXLFs in the two luminosity ranges have 
been calculated for each of 10000 bootstrap-resampled samples. 
As a result, the probability that the ratio reverses, i.e., 
absorbed / unabsorbed ratio becomes greater in the higher luminosity 
bin, turns out to be only less than 0.4\%.

In order to compare absorbed and unabsorbed AGN HXLFs more
quantitatively, we also made a maximum likelihood (ML) technique.
The ML-fitting of the HXLF to the data where individual AGNs have
different spectra is not straightforward. We use U03's approach    
in fitting $N_{\rm H}$ function and the intrinsic HXLF. There is
a limitation in this approach, however, that we have to assume a 
simple absorbed power-law spectrum for each source.
For the fits, we have used the sources in the range 
$z < 0.2$ and $\log\,N_{\rm H} < 24$ to compare with the U03 results.
We again divided into unabsorbed and absorbed AGNs at 
$\log\,N_{\rm H} = 21.5$ in this sample.
In this analysis, the $N_{\rm H}$ function was assumed to be 
constant within each of  $\log\,N_{\rm H} < 21.5$
and $21.5 < \log\,N_{\rm H} < 24$ bins. The fits do not depend 
significantly on details of the evolution within the small
redshift range. Thus, a  pure density evolution model by U03 
($e(z) = (1+z)^{4.2}$) was 
used to represent the small amount of evolution.
The best-fit parameters of the XLFs were calculated following
the same procedure as U03.

Table~\ref{table:5} shows the best fit parameters from the ML-fits.
The fit result from U03's PDE model is also listed. 
The two power-law indices for the 'All' sample are consistent with 
that in U03, while the normalization $A$ seems larger than the U03
value.  This is partially because $A$ is strongly depend on the 
$\log\,L_*$, parameter, for which we found a somewhat smaller value. 
Also, U03 is based on the global fit to all redshifts, where the higher
redshift samples contribute much of the constraints to the 
relevant parameters.
In comparing absorbed and unabsorbed AGN samples, the ML fits have
been made with a fixed   $\log\,L_*=44.0$ in order to avoid strong
dependence between fitting parameters. This value is consistent
with the best-fit values for both samples, when 
fitted with this parameter free.   
As expected, the $\gamma_2$ of absorbed sample is significantly
steeper than that of unabsorbed AGNs.

In Fig.~\ref{HXLF}
(evolution-corrected to $z=0.1$), we overplot the best-fit result of 
smoothed two power law formula for SXLF in the $0<z<0.2$ bin by  
\citet{Hasinger2005}. The $0.5-2$~keV SXLF has been converted to 
$2 - 10$~keV band using a $\Gamma = 1.8$ power-law
spectrum. Our evolution corrected HXLF of unabsorbed AGNs obtained from our 
sample is marginally consistent with the \citet{Hasinger2005} best-fit model.
We also overplot the results from \citet{Sazonov2004} for unasborbed and total
AGNs. They used the $3-20$~keV band, where the attenuation of flux by 
absorption have only minor effect (by $\sim$ 8\% for $\log\,N_{\rm H}=$22.5).
We have thus converted their $3-20$~keV HXLFs (for luminosities 
uncorrected for absorption) to $2-10$~keV using a $\Gamma = 1.8$ power-law 
for overplots. Detailed comparisons are discussed in Sect.~\ref{sec:Disc}.

\section{DISCUSSION}
\label{sec:Disc}

\subsection{Comparisons of Luminosity Functions and Volume Emissivity}

We have calculated the local HXLF with binned $\Sigma V_{\rm a}^{-1}$ method
for our sample defined with {\it HEAO-1} experiment. The HXLF for 
our $\log\,N_{\rm H}<21.5$ results is marginally consistent with the 
best-fit model of SXLF in the $0<z<0.2$ bin in \citet{Hasinger2005} 
as well as with the results from \citet{Sazonov2004} for unasborbed 
and total AGNs. 
It is not surprising that we have small discrepancies because our 
$\log\,N_{\rm H}<$ 21.5 selection does not necessarily match with 
Hasinger's soft X-ray type~1 selection.
A notable exception is the $44 < \log\,L_{\rm x} <44.5$ bin,
where our sample gives an excess which cannot be explained by statistics
alone. 

 The local X-ray volume emissivity of brightest AGNs can be
estimated  by $\Sigma_i (L_{\rm x,i} / V^{i}_{\rm a})$.
The volume emissivity value estimated from our sample is
$\rho_{2-10~{\rm keV}} = 5.85 \pm 1.17 \times 10^{38}$ 
erg s$^{-1}$ Mpc$^{-3}$, 
which is derived using observed luminosity (absorbed).
Note that there is no AGN with $\log\,L_{\rm x} < 42$ erg s$^{-1}$ 
in our sample.

 \citet{Miyaji1994} estimated the $2-10$~keV local volume emissivity 
by modeling the cross-correlation between the all-sky cosmic X-ray 
background (CXRB) surface brightness from the {\it HEAO-1} A-2 
experiment and the {\it IRAS} galaxies, obtaining 
($6.1\pm 1.7 )\times 10^{38}$ ~erg s$^{-1}$ Mpc$^{-3}$
(for $H_0 =$ 70 km s$^{-1}$ Mpc$^{-1}$). This  is consistent with 
the AGN volume emissivity from our sample, implying that the contribution 
of lower luminosity sources  (e.g., star-forming  galaxies and LINERS) 
as well as other sources, non-active galaxies to the local volume 
emissivity is very small. 
On the other hand, \citet{Sazonov2004} estimated the X-ray ($3-20$~keV) 
volume emissivity of $\log\,L_{\rm x} >$ 41 erg s$^{-1}$ from emission 
line AGNs with {\it RXTE} All Sky Slew Survey and found 
$\rho_{3-20~{\rm keV}}$($>~41$) =
($4.9\pm 1.2 )\times 10^{38}$ erg s$^{-1}$ Mpc$^{-3}$, 
(for $H_0 =$ 70 km s$^{-1}$ Mpc$^{-1}$, $z >$ 0.001), which can be converted
to ($2.7\pm 0.7 )\times 10^{38}$ erg s$^{-1}$ Mpc$^{-3}$ 
in the $2-10$~keV band using the absorbed power-law model of AGNs 
estimated based on the {\it RXTE} 
hardness ratios. Their local volume emissivity estimate is only 
about a half of our current estimate. One possibility is an effect of 
local large-scale structure at 
$z<0.015$, where our sample and the {\it RXTE} sample have different 
weights.

% We note that there are 12
%AGNs at $z<0.015$, where local large-scale structure 
%can have a large effect on the local volume emissivity estimates,
%\chg{Their local volume emissivity estimate is only about half of our estimate,%which may be affected by a local large-scale structure
%of $z > 0.015$ AGNs.}
%If we estimate the local volume emissivity only for $z > 0.015$ 
%AGNs in our sample
%(setting $z_{\rm min} = 0.015$ for the $V_{\rm a}$ calculations), 
%we have found $\rho_{2-10~{\rm keV}} = 2.92 \pm 0.59 \times 10^{38}$ 
%erg s$^{-1}$ Mpc$^{-3}$, which now becomes comparable to that in 
%\citet{Sazonov2004}. However, excluding $z<0.015$ AGNs from our sample 
%in effect precludes most AGNs with $\log\,L_{\rm x}\lesssim 43$ 
%from the volume emissivity calculation.

The HXLF from our sample seems larger than that of {\it RXTE} 
with $\log\,L_{\rm x} \sim 44-44.5$. There is a possibility 
of some systematic in \citet{Sazonov2004}'s $2-30$~keV to 
$2-10$~keV conversion. Their absorbed $\Gamma = 1.8$ power-law 
model fit to the {\it RXTE} hardness ratio based on $3-8$ and 
$8-20$~keV count rates is only sensitive to 
$\log\,N_{\rm H}>$ 22. Also their harder band is likely to be affected by 
reflection bumps.   At this moment,
we are not certain whether the differences in the volume emissivity and
HXLF results are due to the large-scale structure (cosmic variance) effect,
assumed spectra in band conversions, or other systematics.
A similar study with a large-scale sensitive hard X-ray survey with 
sufficient AGNs to exclude the $z<0.015$ universe and to sample a large 
volume of the present-day universe, such as available from {\it eROSITA}
or {\it MAXI} will give an ultimate solution.

\subsection{Luminosity --  Absorbed AGN fraction Relation} 

One way of testing the simplest unified scheme, where the 
difference between absorbed and unabsorbed AGNs are purely from
the orientation effect and the intrinsic properties are 
essentially the same is to compare the shapes of unabsorbed 
and absorbed AGN XLFs in {\em intrinsic} (i.e. de-absorbed) luminosity.
 We have found a difference between the HXLFs of the X-ray absorbed 
AGNs and that of unabsorbed AGNs in that the number density of 
absorbed AGNs drops more rapidly than unasborbed ones towards high 
luminosities.  This trend can be also seen in the scatter of the
sample AGNs in the hard X-ray intrinsic luminosity versus 
intrinsic neutral $N_{\rm H}$ diagram in Figure~\ref{RelationBetweenLAndNH},
in which we see a void of AGN in 
$\log\,L_{\rm x} >44.5$, $\log\,N_{\rm H}>$ 21.5. This trend has been 
clearly demonstrated by U03 as the decreasing absorbed fraction of AGNs 
as a function of intrinsic luminosity from a larger sample including 
AGNs in this work, from {\it ASCA} and from {\it Chandra} surveys. 
This lead U03 to suggest a {\em modified} AGN unified scheme, where
the difference between type 1 (unabsorbed) and type 2 (absorbed)
AGNs is not solely due to the viewing angle effect, but involves some
intrinsic difference of the geometric and/or physical conditions 
of the absorbing tori around the SMBH based on the radiative power 
of the primary X-ray component. 

 To verify that the high-luminosity high-absorption void really reflects 
the actual deficiency  of high-luminosity highly absorbed AGNs rather 
than by a selection effect in our sample, we calculate the expected 
number of AGNs in the case where the simplest unified scheme was 
true, i.e., there was no intrinsic difference in the absorbing 
geometry as a function of AGN power. We have started with the smoothed 
two-power-law expression of the $0.5-2$~keV soft X-ray luminosity 
function in the $0.015<z<0.2$ bin  by \citet{Hasinger2005}. We have 
slightly tweaked the parameters and added AGNs in the absorbing columns of 
$\log\,N_{\rm H}$=20, 21., 22, 23, and 24 with the same luminosity function 
shape, allowing the relative normalization to vary,
following an approach similar to an AGN population synthesis
model. The ratios of number densities in each $N_{\rm H}$ bin have been 
adjusted to match the actual numbers of AGNs in $\log\,N_{\rm H}$ 
ranges of $<20.5$, $20.5-21.5$, $21.5-22.5$, $22.5-23.5$, $23.5-24.5$  
respectively. A further  tweak of the parameters have also made to 
reproduce the number of AGNs in $\log\,N_{\rm H}<$ 21.5, 
$\log\,L_{\rm x}> 44.5$ region (luminous unasborbed AGNs). 
An important assumption in this model is that the shape 
(all parameters besides normalization) are identical for 
all $N_{\rm H}$ classes. The expected number of AGNs in the 
``luminous absorbed'' void under this model turns out to be five,
while we observe none. Four of the five AGNs expected under this model 
would be in Sample~1, where only one of 57 X-ray sources are 
unidentified. \citet{Treister2005a} suggested that apparent such effect 
could arise from selection effects, their analysis is based on a 
{\it Chandra} survey focusing on a much fainter population with limited 
identification completeness. Their argument does not apply to our analysis 
or that of U03, involving highly complete samples.  (See also 
\citet{Treister2005b}, where they also found this effect by including
more samples.)

 Recently \citet{Zhang2005} made an interesting suggestion that, 
if most of the intrinsic power-law component in the X-ray spectra of 
AGNs is from a Comptonized emission by optically thick hot plasma above 
the accretion disk, this relation can be explained 
solely by the viewing angle effect due to geometrical projection. 
This assumption needs a physical condition that the planes of the 
accretion disk and the dusty torus are co-aligned. 
We note also that the Comptonizing plasma need to have a planer geometry 
and to be parallel to the accretion disk. If the emission region is 
optically thin and/or composed of hot spots rather than a planer region,
the projection effect would not arise.  

\subsection{Flux Variability}
\label{sec:FV}

\begin{figure}
%\plotone{flux.eps}
\plotone{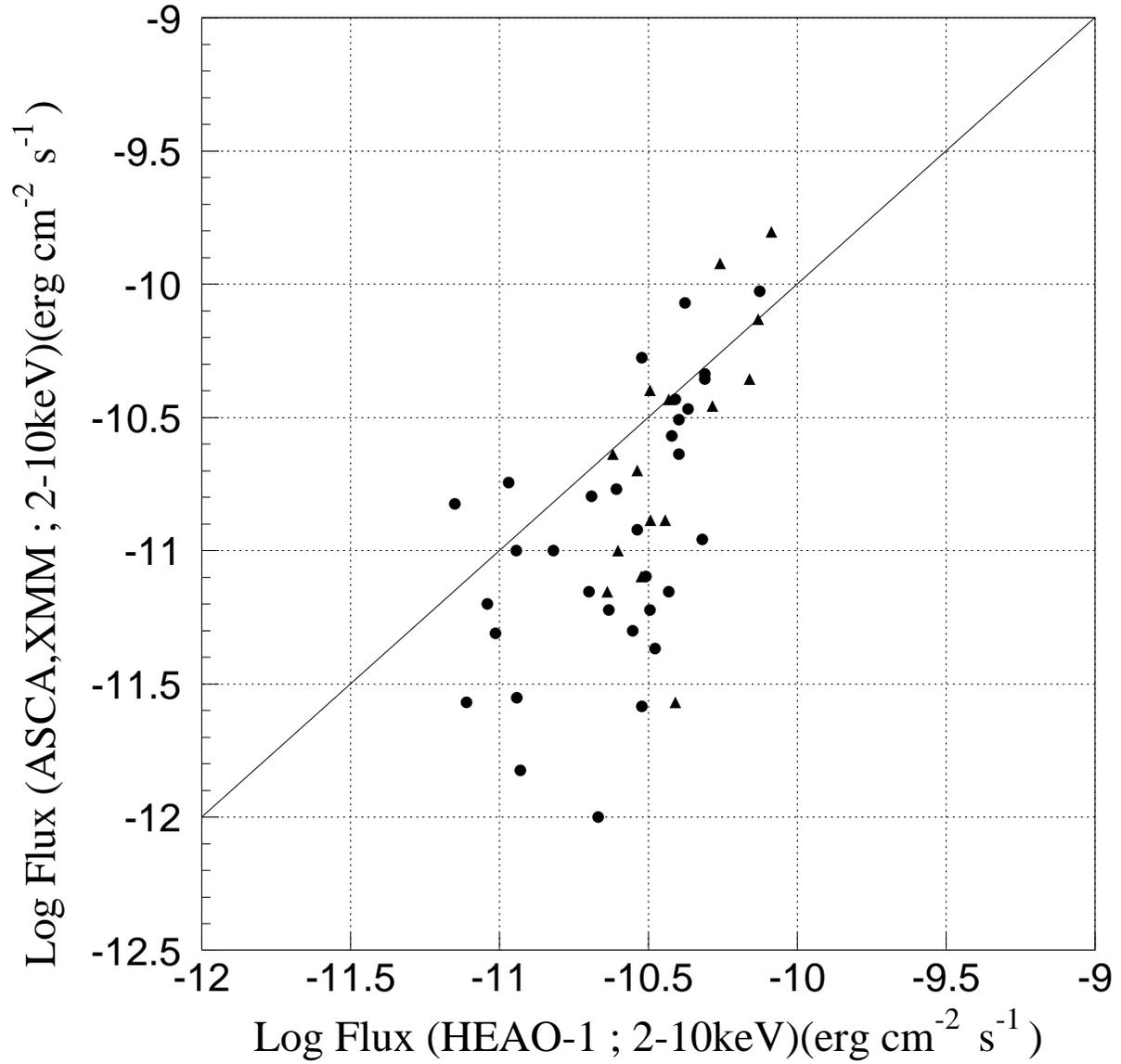}
\caption{Relation between {\it ASCA} or {\it XMM-Newton} observation flux 
and {\it HEAO-1} observation flux. Circle :~unabsorbed AGNs.
Triangle :~absorbed AGNs.
\label{FV}}
\end{figure}

 In our investigation, a complete flux-limited sample of brightest
objects were re-observed with {\it XMM-Newton}, {\it ASCA} and/or 
{\it Beppo-SAX}. We plotted the re-observed flux versus {\it HEAO-1} 
observed flux in Fig.~\ref{FV} for unabsorbed and absorbed AGNs.
The {\it HEAO-1} flux has been corrected for the confusion noise
bias described in Appendix \ref{sec:Conf}. We see from Fig.~\ref{FV} that
most AGNs show lower fluxes upon the re-observations, even after the 
correction.

 The decrease can be naturally explained as a result of AGN variability
and the $\log\,N - \log\,S$ distribution of the AGNs which rises rapidly
as flux becomes lower. 
A flux-limited brightest sample is more likely to pick up an AGN's 
brighter than average state, 
while the re-observation picks up all
states equally. We formulate this effect as follows.

 At sufficiently bright fluxes, the $\log\,N -\log\,S$ relation follows
the Euclidean relation
\begin{equation}
N(>S) \propto S^{-1.5},
\end{equation}
or in the differential form
\begin{equation}
n(S) = d N(S)/d S \propto S^{-2.5}
\label{eq:DiffNS}
\end{equation}
For the convenience of the calculations, we  define $l\equiv \ln\,S$.
Then, Eq.~\ref{eq:DiffNS} can be rewritten as: 
\begin{equation}
 n(l) dl \propto e^{-1.5 l} dl
\label{eq:DiffNL}
\end{equation}

 Suppose that the variability of each individual 
AGN is characterized by the Gaussian log-flux distribution
with a standard deviation $\sigma$:
\begin{equation}
v(l^{\prime},l)=(2\pi\sigma^2)^{-1/2}\exp[-(l-l^{\prime})^2/(2 \sigma^2)],
\label{eq:FluxDist}
\end{equation}
the apparent source count above a log flux of $l_0$ at the time
of the survey observation can be expressed by    
\begin{eqnarray}
N^{\rm obs}(> l_0) = \int ^{\infty}_{-\infty}
\big{\{} n(l^{\prime})\int^{\infty}_{l_0} v(l^{\prime}, l)dl \big{\}} 
dl^{\prime},
\label{eq:FLUX2}
\end{eqnarray}

%$BJ?6Q(Bflux~$l'$$B$NE7BN$,(B$l$$B$H4QB,$5$l$?>l9g$K!"<!$K:F4QB,$7$?;~$N(B
%flux$B$N4|BTCM$O(B$l'$$B$H9M$($i$l$k!#$3$l$h$j!"F1%5%s%W%k$r:F4QB,$7$?>l9g$K(B
%$BF@$i$l$k(Bflu$B8;>/NL$N4|BTCM$O(B
 If an AGN whose average log flux is $l^{\prime}$ was observed at 
$l$ at the time of  the survey, the expectation value of its flux upon 
the re-observation is $l^{\prime}$. Therefore the expectation value
of the {\em decrease} of the flux 
($\langle d l \rangle \equiv \langle l^{\prime} - l \rangle$) 
can be expressed by    
\begin{eqnarray}
\langle dl \rangle = 
\frac{\int ^{\infty}_{-\infty}\big\{n(l^{\prime})
\int ^{\infty}_{l_0}v(l^{\prime}:l)
(l-l^{\prime})dl \big\}dl^{\prime}}
{\int ^{\infty}_{-\infty}\big\{n(l^{\prime})
\int ^{\infty}_{l_0}v(l^{\prime}:l)dl \big\}dl^{\prime}}
\label{eq:FLUX4}
\end{eqnarray}

\begin{figure}
%\plotone{plot-sigma-vs-dl-v2.ps}
\plotone{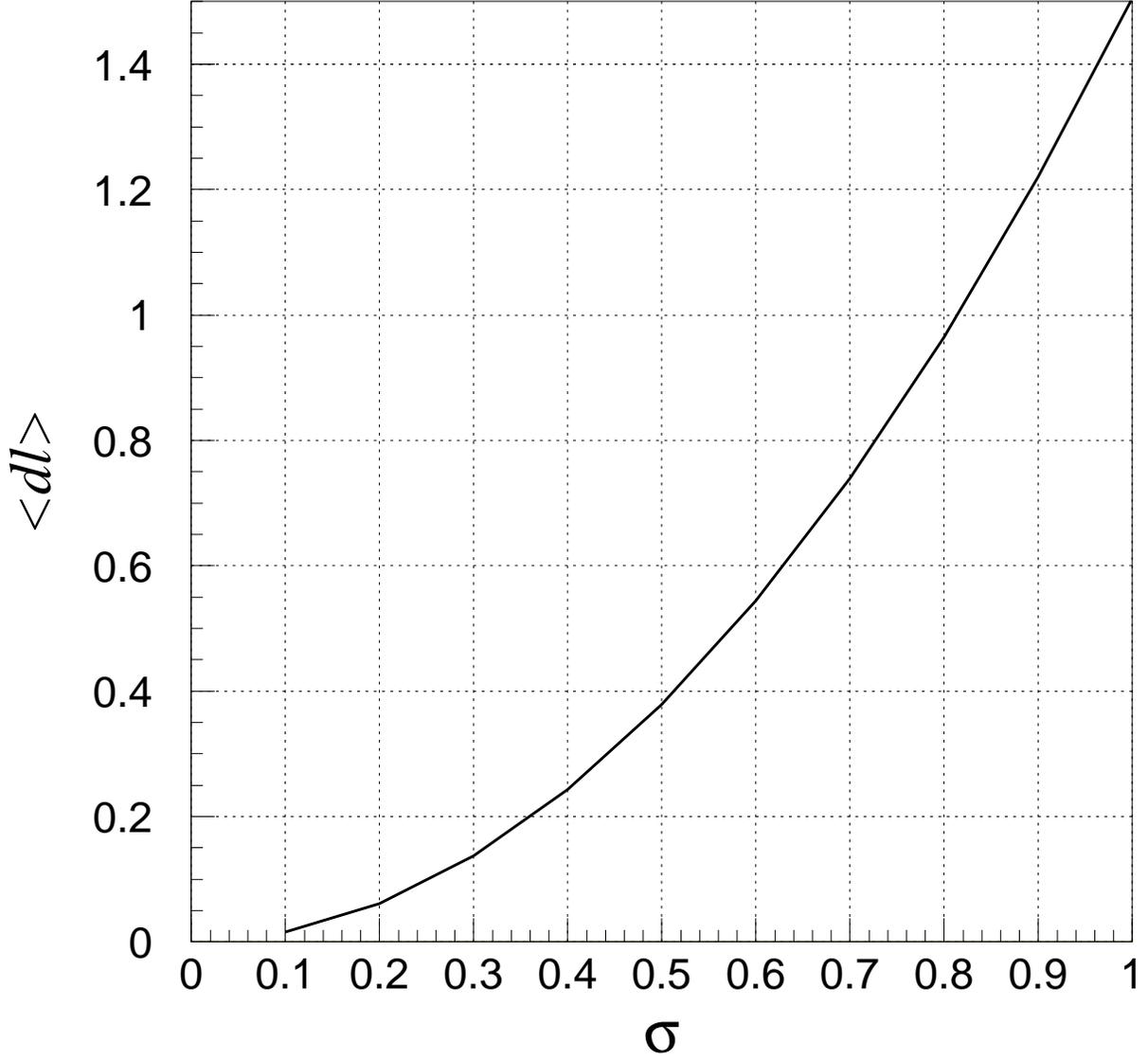}
\caption{The numerical solution of Eq.~\ref{eq:FLUX4} in the 
$\sigma - \langle dl \rangle$ plane, where
$\sigma$ is typical deviation of the variability of AGN flux, 
and $\langle d l \rangle \equiv \langle l^{\prime} - l \rangle$
represents the expectation value of the {\em decrease} of the flux.
We can estimate a typical $\sigma$ of our sample
from this $\sigma$ - $\langle d l \rangle$ relation 
under the assumption discussed in Sect.~\ref{sec:FV}.
\label{fig:Sigma_dl}}
\end{figure}

 Under the assumption that $n(l^{\prime})$ follows the Eucleadean relation 
Eq.~\ref{eq:DiffNL} and log normal flux distribution of AGN flux variation
in Eq. \ref{eq:FluxDist}, and by further assuming that all AGNs vary 
at a typical amplitude characterized by $\sigma$, Eq. \ref{eq:FLUX4} can 
be integrated numerically. Figure \ref{fig:Sigma_dl} shows the numerical 
solution of Eq. \ref{eq:FLUX4} in the $\sigma - \langle dl \rangle$ plane.
>From Fig. \ref{fig:Sigma_dl}, we see that $\langle l \rangle$ is always
positive, which means re-observation give lower fluxes on average.
This is partially because there are much more fainter AGNs under
the distribution Eq.~\ref{eq:DiffNS}.
Thus the net effect of AGN variability is that AGN in the
flux-limited sample seems to show a flux decrease upon the re-observation 
on average. In turn, from the average decrease of the flux, we can estimate 
the typical AGN variability amplitude. In our sample, the re-observed 
flux by {\it ASCA}, {\it XMM-Newton} or Beppo-SAX was smaller than 
the {\it HEAO-1} flux by a factor of 0.28 on average (after correction 
for spectral response and confusion). Figure \ref{fig:Sigma_dl} shows that 
this corresponds to $\sigma =$ 0.914 or a typical variability amplitude of 
a factor of 2.5.

%equation~\ref{eq:FLUX3}$B5Z$S(B~\ref{eq:FLUX4}$B$O?tCM7W;;$G=P$9;v$,$G$-$k!#(B
%$B$3$l$h$j(B

\section{CONCLUSIONS AND PROSPECTS}

 We summarize our investigation and main conclusions below.

\begin{enumerate}
\item We have defined a complete flux-limited sample from a 
 combination of {\it HEAO-1} A2 and A1/A3 surveys with a 
 well-defined flux-survey area relation. The sample
 is composed of 28 AGNs from \citet{Piccinotti1982} 
 and 21 AGNs from a deeper sample by \citet{Grossan1992}.

\item A bias to the measured flux due to confusion noise
 have been found to be significant in Sample~2 (A1/A3). 
 We have modeled and made first-order corrections to 
 the fluxes and survey areas. 

\item We have obtained X-ray spectral information for all (but one)
 AGNs in the combined sample from {\it XMM-Newton}, {\it ASCA}
 and {\it Beppo-SAX}.

\item The spectra have been first modeled as an absorbed 
 power-law. We have made closer look at 6 AGNs which gave 
 spectral index of  $\Gamma<1.4$ in this fit and used models
 involving (multiple) partial covering and/or ionized absorbers.

\item We have found the mean photon spectral index of 
  $\Gamma \approx 1.76$ with a 1$\sigma$ dispersion of 0.2.

\item We find the $N_{\rm H}$ distribution for our sample,
 which can be used as a constraint to X-ray population 
 synthesis modeling.

\item We have constructed local hard X-ray luminosity 
 functions (HXLFs) from our sample, separately for absorbed and
 unabsorbed AGNs, as well as for both, as a function of the 
 {\em intrinsic} $2-10$~keV luminosity. We have also made  
 HXLFs which have been corrected for density evolution.
 
\item The scatter of the sample AGNs in the  $N_{\rm H}$ versus
 intrinsic luminosity plane shows a void of AGNs at a 
 high-luminosity, high absorption regime, where we expect
 $\sim 5$ AGNs if the intrinsic luminosity functions between
 absorbed and unabsorbed AGNs were identical. This, as well 
 as the difference in the HXLFs between absorbed and unabsorbed
 AGNs, is not likely due to selection effects.  

\item  The X-ray fluxes of AGNs in the sample observed 
 by {\it XMM-Newton}, {\it ASCA} and/or {\it Beppo-SAX} were 
 on average lower than those observed by {\it HEAO-1} for the
 same objects. The mean flux decrease upon re-observation of 
 AGNs in a flux-limited sample has been formulated can be naturally
 explained by the AGN variability, where the AGNs are likely to have
 brighter-than-average flux at the time of the sample-defining survey.
\end{enumerate}

 As a concluding remark, we note that the sensitivity of large-area 
surveys at hard X-ray bands currently available is still not 
sufficient, in contrast with the soft band, where {\it ROSAT} 
All-Sky Survey produced enormous dataset to sample the 
present-day universe unabsorbed AGN population. In this work, 
we had to depend on data obtained in the 1970's to define our 
sample, to include the absorbed AGN population. 
Because of this, our investigation is subject to fundamental limitations 
in terms of object count statistics, confusion and mis-identification 
problems. To obtain sufficient number of AGNs,
we also had to use very nearby objects ($z<0.015$), where 
the quantities such as HXLF or volume emissivity is subject to
the density fluctuations due to the local large scale structure.    
The understanding of AGN evolution is never complete until we 
sample the present-day universe fairly. 
 In the near future, a slew survey from the  {\it Swift} BAT 
mission \citep{Markwardt2005} and a dedicated 
all-sky exposure with {\it INTEGRAL} will produced a less biased 
bright end catalog of AGNs and after a spectroscopic followup
such as this work, we expect improvements in our knowledge in this
region.  Future missions such as  {\it MAXI}, and {\it eROSITA} would find 
numerous AGNs in our local universe will enable us to make a precision 
statistics of AGNs in the present-day universe.

\acknowledgments

We thank the anonymous referee for his careful review and 
useful suggestions.
This work has been supported by NASA Grants NAG5-10043
({\it XMM-Newton} Genral Observer) and NAG5-10875 (LTSA) to TM.
Part of this work was financially supported
by Grant-in-Aid for Scientific
Research (No.\ 16340077) from
the Japan Society for the Promotion of Science.
This research has made use of data obtained from the High Energy 
Astrophysics Science Archive Research Center (HEASARC),
provided by NASA's Goddard Space Flight Center.

\clearpage

\appendix
\section{Correcting for Confusion Effects on {\it HEAO-1} A1 Counts}
\label{sec:Conf}

\begin{figure}
%\plotone{xlf_1_cfw.eps}
\plotone{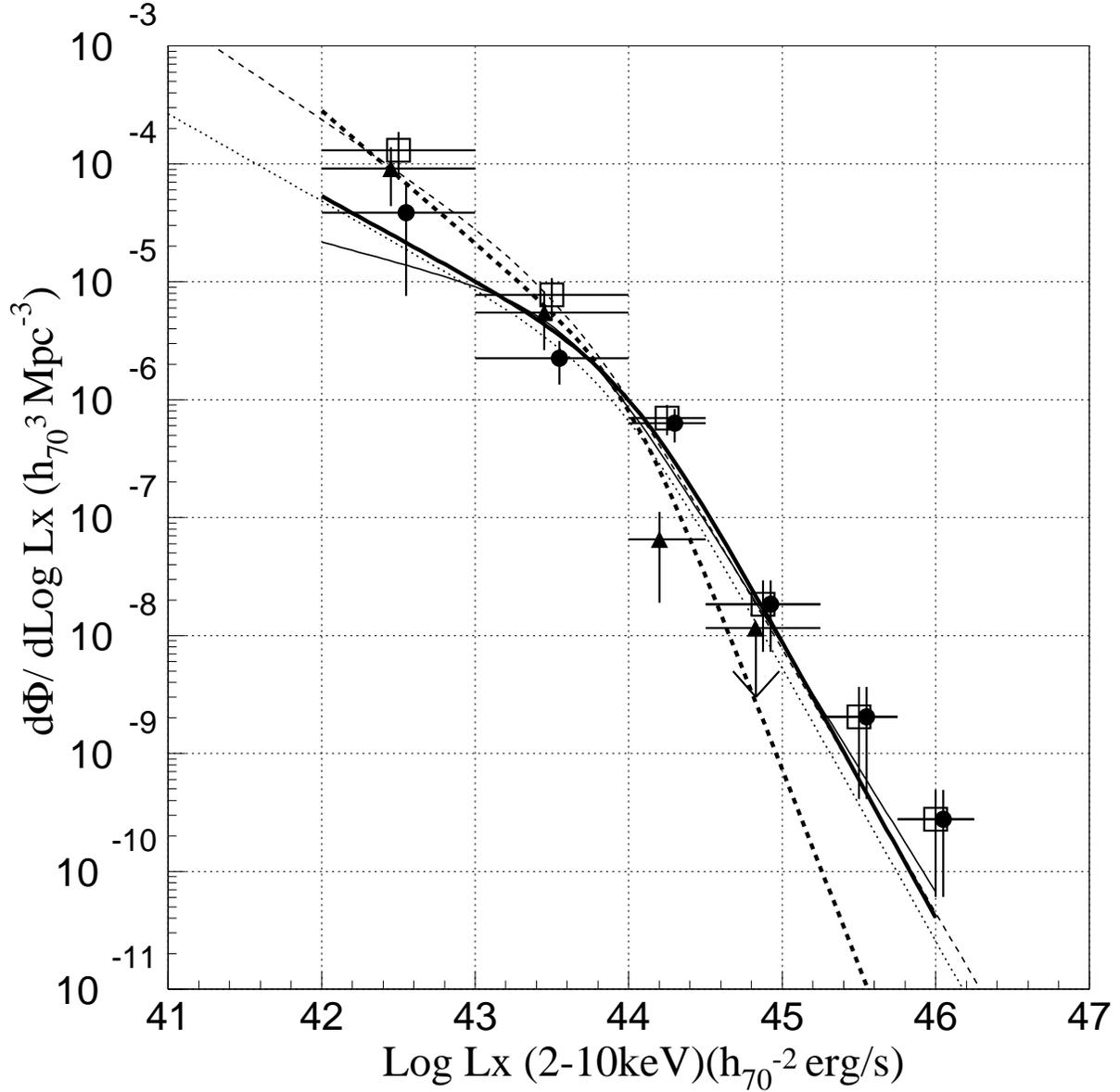}
\caption{Hard X-ray luminosity functions for all AGNs (black, square), X-ray
unabsorbed AGNs (red, circle), and absorbed AGNs (blue, triangle) in 
our sample for the evolution-corrected case. 
Thick solid lines and dotted lines show 
the ML-fitting results of the XLFs for unabsorbed and absorbed
AGNs, respectively.
Thin solid lines show type 1 AGN XLF from \citet{Hasinger2005}. 
Thin dashed and dotted 
lines show the results from \citet{Sazonov2004} for all AGNs and unabsorbed
AGNs respectively. All overplotted lines have been converted to 2-10 keV
from the originals assuming a $\Gamma = 1.8$ power-law.
Errors are 1$\sigma$.
\label{HXLF}}
\end{figure}

 Our sample selections are partially based on count rates (CRs) observed
by the {\it HEAO-1} A1 experiment, which have been cataloged by 
\citet{Wood1984} (hereafter W84). The CR limit imposed by 
\citet{Grossan1992} of $CR_{\rm lim}=3.6\times 10^{-3}$ LASS 
cts s$^{-1}$ cm$^{-2}$ is subject to confusion noise, in addition to the 
formal error given by W84, which only includes the effects of 
photon counting statistics. Due to these effects, the CRs are subject 
to systematic overestimations near the detection limit. This may bias 
the estimates of luminosity function derived in Sect.~\ref{HXLF}.
%
% Give a few numbers
%

 We have made a first-order correction to these effects by 
Monte-Carlo simulations.
We have taken the  following steps.

\begin{figure}
%\plotone{a1sim1_plot4_bw.eps}
\plotone{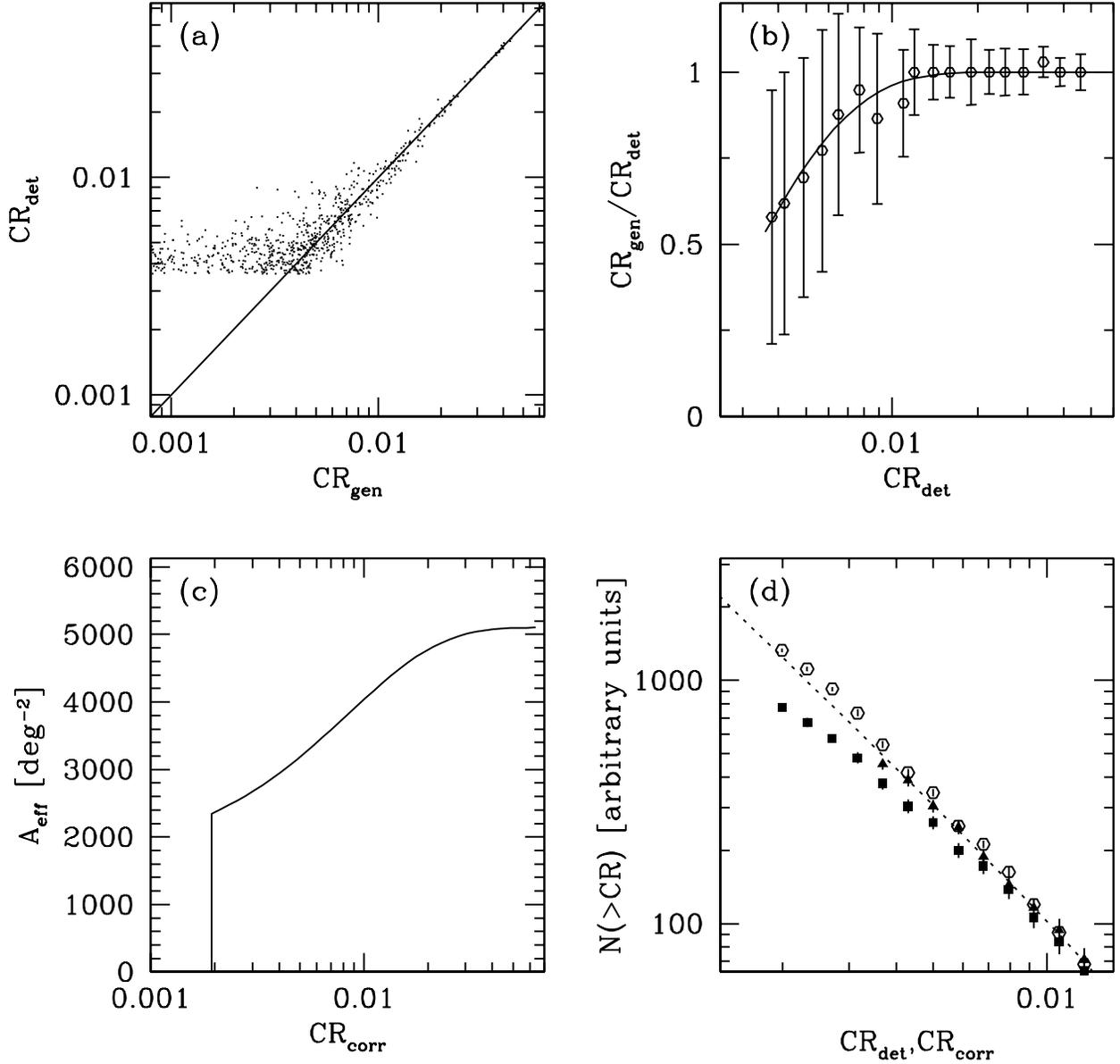}
\caption{(a) The scatter diagram of the generated ``true'' count rate 
  ($CR_{\rm gen}$) versus the detected count rate ($CR_{\rm det}$) in 
  our simulation. The solid line shows $CR_{\rm gen}=CR_{\rm det}$. 
  (b) The $\langle CR_{\rm gen} \rangle/CR_{\rm det}$ value is plotted as 
  a function of $CR_{\rm det}$. The error bars indicate the 1$\sigma$ 
  dispersion for the simulated object in the $CR_{\rm det}$ bin and 
  {\em not} the standard deviation of the mean. The solid curve shows 
  the fit to Eq. \ref{eq:cr_corr}. (c) The corrected area curve for 
  Sample~2 from Eqs. \ref{eq:effarea} \& \ref{eq:arcorr_para}. 
  (d) The $\log N - \log S$ curves for the simulations. The dashed line 
  shows the input model.  Solid  triangles show the 
  $\log N - \log S$ calculated from $CR_{\rm det}$ with the 
  geometrical area. Sold squares show the case $CR_{\rm corr}$ is used with 
  the geometrical area. Open squares show the case after 
  correcting both CR and the area ($CR_{\rm corr}$ used with $A_{\rm eff}$).
\label{fig:crsim}}
\end{figure}

\begin{enumerate}
\item Sources are generated based on approximate $\log N - \log S$ relation, 
  extrapolated down to approximately one order of magnitude below our 
  flux limit. In this simulation, LASS CRs ($CR_{\rm A1}$) have been used  
  rather than fluxes in physical units, because we define our limit
  by CRs.  From the W84 catalog, we have estimated: 
  \begin{equation}
    N(>CR_{\rm A1}) = 7\times 10^{-3} (CR_{\rm A1}/5\times 10^{-3})^{-1.5} 
      [{\rm deg}^{-2}] 
      \label{eq:a1ns}
  \end{equation}

\item For each generated source with true underlying CR
   ($CR_{\rm gen}$), an observed CR ($CR_{\rm obs}$)  has been simulated
   by
  \begin{equation}
    CR_{\rm obs} = CR_{\rm gen} + \Delta CR_{\rm err} + \Delta CR_{\rm cfs},
  \end{equation}
    where $\Delta CR_{\rm err}$ simulates the deviation due to photon counting statistics
    and $\Delta CR_{\rm cfs}$ due to confusion noise as detailed below. 
    \begin{itemize}
      \item For a given $CR_{\rm gen}$, an error $\sigma$ is selected from
         objects in $|b|>30$, $|EB|>35$, and $CR<0.01$, 
         where EB is the ecliptic latitude. $\Delta CR_{\rm err}$
         is derived from a random Gaussian deviation using the selected $\sigma$.
         This may underestimate errors for brighter sources, but the effects we consider
         is only non-negligible for sources near detection limits.
      \item  $\Delta CR_{\rm cfs}$ is derived from Monte-Carlo simulations, where
         objects are generated based on Eq. \ref{eq:a1ns} in the range 
         $5\times 10^{-5}<CR<3.6\times 10^{-3}$ (the upper bound is our flux 
         limit) and distributed over the {\it HEAO-1} A1 beam (W84). The total 
         CRs folded with the beam ($CR_{\rm beam}$) have been calculated for 
         2000 runs. If $CR_{\rm beam}$ is greater than 
         3.6$\times$10$^{-3}$ above the mean value, the run is rejected.
     \item The mean value ($\langle CR_{\rm beam} \rangle$) has been recalculated 
        after the rejection. Then $\Delta CR_{\rm cfs}$ has been calculated as
        $CR_{\rm beam} -  \langle CR_{\rm beam}\rangle$, where $CR_{\rm beam}$
        is taken from one of the simulation runs which was not rejected.
    \end{itemize} 

\item  In order to simulate detected sources, only those with 
    $CR_{\rm obs}>3.6\times 10^{-3}$ have been selected  ($CR_{\rm obs}$ 
    of the selected objects is indicated by $CR_{\rm det}$, representing
    'detected count rates').  Figure \ref{fig:crsim}(a) shows the scatter diagram 
    of $CR_{\rm det}$  versus  $CR_{\rm gen}$ for those objects. The simulated 
    objects have been then binned by  $CR_{\rm det}$ and the mean value of 
    $CR_{\rm gen}$ ($\langle CR_{\rm gen}\rangle$) has been calculated in  
    each bin. This  gives a measure of mean expected true CR of the sources 
    as a function of the detected CR.

\item Our next step is to find a smooth analytical representation of 
     $\langle CR_{\rm gen} \rangle/CR_{\rm det}$ 
     as a function of $CR_{\rm obs}$. Since we know that  
     $\langle CR_{\rm gen} \rangle/CR_{\rm det}$ 
     goes asymptotically close to unity at higher count rates, we have used 
     the form:
     \begin{eqnarray}
       C_{\rm cr}\equiv \langle CR_{\rm det} \rangle/CR_{\rm det} 
       &= 1-\exp[-(CR_{\rm det}-s_{\rm cr})/w_{\rm cr})], \nonumber\\ 
       & (CR_{\rm det}\geq CR_{\rm})
       \label{eq:cr_corr}
     \end{eqnarray}  
     where $C_{\rm cr}$ represents the CR correction factor.
     By a $\chi^2$ fit with the standard deviation of $CR_{\rm gen}$
     in individual bins as $\sigma$, we find the best-fit values: 
     \begin{equation}
       s_{\rm cr} =1.6\times 10^{-3}; w_{\rm cr}=2.6\times 10^{-3}.         
      \label{eq:crcorr_para}
    \end{equation}
     The binned $\langle CR_{\rm gen} \rangle/CR_{\rm det}$ values
     are plotted against $CR_{\rm det}$ in Fig. \ref{fig:crsim}(b) along 
     with the best-fit function form.  
     Note that we use $\chi^2$ with these $\sigma$'s only as a weighting scheme 
     for the fit and the formal parameter errors of the $\chi^2$ statistics 
     does not have any meaning. 

\item The detected CRs have been corrected by $CR_{\rm corr} = C_{\rm cr} ~CR_{\rm det}$.
     By this correction, the estimated 'true' count rate for a source detected 
     at the faintest limit of $CR_{\rm det}=CR_{\rm lim}=3.6\times 10^{-3}$ 
     becomes $CR_{\rm corr,lim}=1.9\times 10^{-3}$.   

\item In calculating the XLF, we also have to correct the survey area for  
     incompleteness of the detection, where not all sources with a true count rate of, 
     e.g. $>1.9\times 10^{-3}$ will have $CR_{\rm obs}>3.6\times 10^{-3}$.
     This is made by dividing the  differential $\log N - \log S$ 
     curves of $CR_{\rm corr}$ by that of $CR_{\rm gen}$ calculated using the simple 
     geometrical area ($A_{\rm geo}$) of the survey. Using the same functional form as  
     Eq. \ref{eq:cr_corr}, the {\em effective} survey area as a function of
     $CR_{\rm corr}$ is expressed by
     \begin{eqnarray}
       A_{\rm eff}/A_{\rm geo} & = 1-\exp[-(CR_{\rm corr}-s_{\rm A})/w_{\rm A}], & 
                                       (CR_{\rm corr}\geq CR_{\rm corr,lim}) \nonumber\\
                               & = 0,  & (CR_{\rm corr}<CR_{\rm corr,lim}).                           
       \label{eq:effarea}
     \end{eqnarray}  
     We have found the best-fit values of the parameters:
     \begin{equation}
          s_{\rm A} = -3.3\times 10^{-3};  w_{\rm A}=8.5\times 10^{-3}.         
             \label{eq:arcorr_para} 
     \end{equation}         
     The resulting effective area curve is plotted against $CR_{\rm corr}$ in 
     Fig \ref{fig:crsim}(c).

\item Using the corrected $CR_{\rm corr}$ and $A_{\rm eff}$, a corrected 
     $\log N - \log S$ curve has been calculated to see that it is in reasonable 
     agreement with the initially assumed one. This curve is plotted 
     in  Fig \ref{fig:crsim}(c) along with those from  $CR_{\rm det}$ and 
      $CR_{\rm corr}$ using the geometrical area. 

\item Now we are ready to calculate the luminosity function from the
     real data by applying the corrections Eqs. \ref{eq:cr_corr} \& \ref{eq:effarea}. 
     to the cataloged CR and geometrical area respectively. 
\end{enumerate}

 We have also made the same experiment for Sample~1 
based on the {\it HEAO-1} A2 \citet{Piccinotti1982} catalog. 
The effects are negligible in the case of Sample~1.

%======== TMEDIT END ==========

\clearpage
% ==================================================

%%%%%%%%%%%  References  %%%%%%%%%%%%%%%%%%%%%%%%%%%

%%\bibliography{hxcs0.7}   % Produces the bibliography via BibTeX.

%%%%%%%%%%%  Figure %%%%%%%%%%%%%%%%%%%%%%%%%%%%%%%%%


%% ================ TMEDIT ==================

%%%%%%%%%%%  Table %%%%%%%%%%%%%%%%%%%%%%%%%%%%%%%%%

\begin{deluxetable}{clcccccccc}
\tabletypesize{\scriptsize}
\tablecaption{Source List~--~Sample~1 \label{table:1-1}}
\tablewidth{0pt}
\tableheadfrac{0.1}
\tablehead{
\colhead{Source ID} & \colhead{Name} & \colhead{RA}   & \colhead{Dec}  & \colhead{Type\tablenotemark{a}} &
\colhead{Redshift}  & \colhead{Galactic N$_{\rm H}$}\\
&&\colhead{(J2000.0)}&\colhead{(J2000.0)}&&($z$)&\colhead{(10$^{20}$ cm$^{-2}$)}}
\startdata
1  & IC 4329A         & 13 49 19.2 & -30 18 34  & S1    & 0.0168  & 4.42   \\
2  & 3C 273           & 12 29 06.7 &  02 03 08  & QSO   & 0.158   & 1.79   \\
3  & NGC 2992         & 09 45 41.9 & -14 19 15  & S1.9  & 0.0076  & 5.37   \\
4  & NGC 5506         & 14 13 14.9 & -03 12 27  & S1.9  & 0.0072  & 3.81   \\
5  & NGC 526A         & 01 23 54.8 & -35 03 55  & S1.5  & 0.0202  & 2.20   \\
6  & NGC 7582         & 23 18 23.0 & -42 22 11  & S2    & 0.0044  & 1.59   \\
7  & ESO 198-G024     & 02 38 19.6 & -52 11 34  & S1    & 0.045   & 3.13   \\
8  & MCG -6-30-15     & 13 35 53.3 & -34 17 48  & S1    & 0.0082  & 4.08   \\
9  & MKN 509          & 20 44 09.0 & -10 43 15  & S1    & 0.0352  & 4.11   \\
10 & 3C 120           & 04 33 11.0 &  05 21 15  & S1    & 0.033   & 11.1   \\
11 & NGC 7172         & 22 02 01.7 & -31 52 18  & S2    & 0.0085  & 1.65   \\
12 & NGC 3783         & 11 39 02.0 & -37 44 19  & S1    & 0.0096  & 8.50   \\
13 & MKN 926          & 23 04 43.4 & -08 41 08  & S1.5  & 0.047   & 3.60   \\
14 & NGC 7469         & 23 03 16.0 &  08 52 26  & S1    & 0.0161  & 4.87   \\
15 & NGC 4593         & 12 39 39.0 & -05 20 39  & S1    & 0.0099  & 2.31   \\
16 & NGC 3227         & 10 23 30.5 &  19 51 55  & S1.5  & 0.0052  & 2.36   \\
17 & ESO 141-G55      & 19 21 14.2 & -58 40 12  & S1    & 0.0371  & 5.09   \\
18 & IIIZw2           & 00 10 31.0 &  10 58 30  & S1    & 0.0898  & 5.72   \\
19 & Fairall 49       & 18 36 58.4 & -59 24 07  & S2    & 0.020   & 7.13   \\
20 & 3C 445           & 22 23 49.0 & -02 06 12  & S1    & 0.0562  & 5.01   \\
21 & MKN 1152         & 01 13 50.1 & -14 50 44  & S1.5  & 0.0527  & 1.69   \\
22 & NGC 7314         & 22 35 46.0 & -26 03 02  & S1.9  & 0.0046  & 1.46   \\
23 & MCG -1-24-12     & 09 20 46.3 & -08 03 22  & S2    & 0.0198  & 3.56   \\
24 & ESO 103-G35      & 18 38 20.3 & -65 25 42  & S1.9  & 0.0133  & 7.64   \\
25 & 3A 0557-383      & 05 58 02.1 & -38 20 05  & S1    & 0.0344  & 3.98   \\
26 & MRK 590          & 02 14 33.6 &  00 46 00  & S1    & 0.0263  & 2.68   \\
27 & H 1846-786       & 18 47 02.8 & -78 31 50  & S1    & 0.076   & 9.05   \\
28 & Fairall 9        & 01 23 46.0 & -58 48 21  & S1    & 0.046   & 3.19   \\
\enddata

\tablenotetext{a}{S1~:~Seyfert 1,~~S2~:~Seyfert 2 ,~~S1.5~:~Seyfert 1.5,~~S1.9~:~Seyfert 1.9}

%%\tablecomments{Occasionally, authors wish to append a short
%%paragraph of explanatory notes that pertain to the entire table, but
%%which are different than the caption.  Such notes should be placed in
%%a {\tt tablecomments} command like this.}
\end{deluxetable}
%Source List~--~Sample~1
\begin{deluxetable}{clccccc}
\tabletypesize{\scriptsize}
\tablecaption{Source List~--~Sample~2 \label{table:1-2}}
\tablewidth{0pt}
\tableheadfrac{0.5}
\tablehead{
\colhead{Source ID} & \colhead{Name} & \colhead{RA}   & \colhead{Dec}  & \colhead{Type\tablenotemark{a}} &
\colhead{Redshift}  & \colhead{Galactic N$_{\rm H}$}\\
&&\colhead{(J2000.0)}&\colhead{(J2000.0)}&&($z$)&\colhead{(10$^{20}$ cm$^{-2}$)}}

\startdata
29  & Kaz 102          & 18 03 28.8 & 67 38 10  & S1      & 0.136  &  4.62       \\
30  & KUV 18217+6419   & 18 21 57.3 & 64 20 36  & S1      & 0.297  &  4.05       \\
31  & MKN 885          & 16 29 48.3 & 67 22 42  & S1.5    & 0.026  &  3.85       \\
32  & MKN 876          & 16 13 57.2 & 65 43 10  & S1      & 0.129  &  2.87       \\
33  & 3C 390.3         & 18 42 09.0 & 79 46 17  & S1      & 0.057  &  4.24       \\
34  & MKN 290          & 15 35 52.3 & 57 54 09  & S1      & 0.030  &  1.72       \\
35  & MKN 279          & 13 53 03.4 & 69 18 30  & S1.5    & 0.0297 &  1.78       \\
36  & H 1318+692       & 13 20 24.6 & 69 00 12  & S1      & 0.068  &  1.75       \\
37  & H 1419+480       & 14 21 29.6 & 47 47 27  & S1.5    & 0.072  &  1.65       \\
38  & H 1320+551       & 13 22 49.2 & 54 55 29  & S1      & 0.064  &  1.36       \\
39  & PG 0804+761      & 08 10 58.5 & 76 02 43  & S1      & 0.100  &  2.97       \\
40  & MKN 506          & 17 22 39.9 & 30 52 53  & S1.5    & 0.043  &  3.26       \\
41  & MRK 6            & 06 52 12.3 & 74 25 37  & S1.5    & 0.0197 &  6.39       \\
42  & H 1537+339\tablenotemark{c} & 15 39 52.2 & 33 49 31  & S1 & 0.329&2.06     \\
43  & MKN 478          & 14 42 07.5 & 35 26 23  & S1      & 0.077  &  1.03       \\
44  & MKN 464          & 13 55 53.5 & 38 34 29  & S1.5    & 0.051  &  0.96       \\
45  & NGC 5033         & 13 13 27.0 & 36 35 39  & LINER?  & 0.0047 &  1.03       \\
46  & PG 1425+267      & 14 27 35.7 & 26 32 14  & QSO     & 0.366  &  1.73       \\
47  & NGC 4151         & 12 10 32.6 & 39 24 21  & S1.5    & 0.0051 &  1.98       \\
48  & AKN 564          & 22 42 39.3 & 29 43 31  & S1.9    & 0.0247 &  6.40       \\
49  & NGC 5548         & 14 17 59.5 & 25 08 12  & S1.5    & 0.0177 &  1.69       \\
\enddata

\tablenotetext{a}{S1~:~Seyfert 1,~~S2~:~Seyfert 2 ,~~S1.5~:~Seyfert 1.5,~~S1.9~:~Seyfert 1.9}
\tablenotetext{c}{No data with ASCA or XMM-Newton observation}

\end{deluxetable}
%Source List~--~Sample~2
\begin{deluxetable}{clccccccc}
\tabletypesize{\scriptsize}
\tablecaption{Results~--~Sample~1 \label{table:2-1}}
\tablewidth{0pt}
\tableheadfrac{0.1}
\tablehead{
\colhead{Source ID} & \colhead{Name} & \colhead{Flux\tablenotemark{a}}&
\colhead{L$_{\rm X}$\tablenotemark{b}}  &
\colhead{$\Gamma$} &\colhead{Intrinsic N$_{\rm H}$\tablenotemark{c}} &
\colhead{Flux\tablenotemark{a}} & \colhead{Obs.(ref)\tablenotemark{d}}
&\colhead{Exp.time\tablenotemark{e}}\\
& & \colhead{HEAO-1} & \colhead{(10$^{43}$~erg/s)} & &
\colhead{(10$^{20}$~cm$^{-2}$)} & \colhead{ASCA-XMM}&&\colhead{(ks)}
}
\startdata
1  & IC 4329A         &  8.2 &   5.6  & 1.741$^{+0.011}_{-0.012}$ & 35.8$^{+0.7}_{-0.7}$ & 15.7 &  XMM  & 12.7/9.90  \\
2  & 3C 273           &  7.5 & 540.   & 1.658$^{+0.029}_{-0.026}$ & 0$^{+9.7}$           & 9.4  &  XMM  & 8.14/5.96  \\
3  & NGC 2992         &  7.4 &   1.0  & 1.7                       & 90                   & 7.4  &  Gi00 & 59.2/27.1   \\
4  & NGC 5506         &  5.5 &   0.80 & 1.721$^{+0.05}_{-0.049}$  &323$^{+19}_{-19}$     & 12.0 &  XMM  & 13.1/9.92  \\
5  & NGC 526A         &  5.2 &   5.3  & 1.61$\pm$0.02&115$\pm$3                          & 3.5  &  T97  & 43.4/51.0  \\
6  & NGC 7582(*)      &  3.9 &   0.86 & 1.95                      &17440                 & 1.97 &  Tu00 & 56.4/52.2   \\
7  & ESO 198-G024     &  4.8 &  23.   & 1.77$\pm$0.03             &$<$3.2                & 1.1  &  Gu03 & 0/6.8      \\
8  & MCG -6-30-15     &  4.9 &   0.73 & 1.92$^{+0.02}_{-0.01}$    &1.7$^{+0.4}_{-0.3}$   & 4.6  &  R97  & 147/       \\
9  & MKN 509          &  4.3 &  13.   & 1.494$^{+0.029}_{-0.028}$ &0$^{+4.8}$            & 3.4  &  XMM  & 24.7/0     \\
10 & 3C 120           &  4.9 &  12.   & 2.00$^{+0.01}_{-0.02}$    &4.6$^{+0.7}_{-0.5}$   & 4.4  &  G98  & 47.4/      \\
11 & NGC 7172         &  3.7 &   0.97 & 1.69$^{+0.08}_{-0.09}$    &819$^{+34.2}_{-33}$   & 3.7  &  T97  & 14.9/15.6  \\
12 & NGC 3783         &  4.2 &   0.87 & 1.60$^{+0.02}_{-0.02}$    &8.7$^{+19}_{-19}$     & 8.5  &  BL02 & 37.7/37.3  \\
13 & MKN 926          &  4.0 &  21.0  & 1.612$^{+0.091}_{-0.071}$ &0$^{+26.3}$           & 3.1  &  XMM  & 10.3/      \\
14 & NGC 7469         &  4.0 &   2.3  & 1.770$^{+0.040}_{-0.042}$ &0$^{+7.3}$            & 2.3  &  XMM  & 17.8/12.3  \\
15 & NGC 4593         &  3.9 &   0.86 & 1.692$^{+0.049}_{-0.05}$  &0$^{+8.8}$            & 3.7  &  XMM  & 13.9/9.5   \\
16 & NGC 3227         &  3.0 &   0.25 & 1.52 \tablenotemark{f}    & 660\tablenotemark{g} & 0.8  &  GO03(1)& 37.4/35.3\\
17 & ESO 141-G55      &  3.8 &  12.   & 1.72$\pm$0.06             & $\leq 1$             & 2.7  &  GO03(2)& 57.0/57.6\\
18 & IIIZw2           &  3.7 &  76.   & 1.75$^{+0.018}_{-0.017}$  &0$^{+0.21}$           & 0.7  &  XMM  & 7.32/10.1  \\
19 & Fairall 49       &  3.6 &   3.5  & 1.983$^{+0.044}_{-0.044}$ &92.2$^{+5.4}_{-5.3}$  & 1.3  &  ASCA & 20.2/34.5  \\
20 & 3C 445(*)        &  2.3 &  28.   & 1.108$^{+0.157}_{-0.118}$ &1321$^{+114}_{-158}$  & 0.7  &  XMM  & 0/15.3     \\
21 & MKN 1152         &  3.2 &  21.   & 1.672$^{+0.049}_{-0.049}$ &13.8$^{+3.4}_{-3.2}$  & 0.6  &  ASCA & 33.8/37.4  \\
22 & NGC 7314         &  3.2 &   0.16 & 1.848$^{+0.013}_{-0.012}$ &72$^{+1}_{-1}$        & 4.0  &  XMM  & 42.3/30.4  \\
23 & MCG -1-24-12     &  2.5 &   3.1  & 1.59$^{+0.07}_{-0.03}$    &627$^{+53}_{-57}$     & 1.0  &  Ma02 & 76/36      \\
24 & ESO 103-G35      &  2.4 &   2.5  & 1.955$^{+0.203}_{-0.193}$ &2028$^{+132}_{-125}$  & 2.3  &  XMM  & 9.7/0      \\
25 & 3A 0557-383(*)   &  2.9 &  10.   & 1.8 \tablenotemark{f}     &37                    & 2.0  &  T96  & 40/40      \\
26 & MRK 590          &  2.8 &   4.5  & 1.526$^{+0.077}_{-0.75}$  &0$^{+12.3}$           & 0.5  &  XMM  & 10.4/7.0   \\
27 & H 1846-786       &  3.1 &  44.   & 1.929$^{+0.025}_{-0.024}$ &1.2$^{+1.5}_{-1.2}$ & 0.8    &  ASCA & 59.9/76.9  \\
28 & Fairall 9        &  2.9 &  15.   & 1.751$^{+0.051}_{-0.044}$ &$\leq 1$             & 1.2   &  XMM  & 28.7/27.7  \\
\enddata

\tablenotetext{a}{Observed After absorption,~(10$^{-11}$~erg~cm$^{-2}$~s$^{-1}$) }
\tablenotetext{b}{Intrinsic X-ray luminosity in $2-10$~keV band (erg~s$^{-1}$)}
\tablenotetext{c}{After subtracted with Galactic N$_{\rm H}$ }
\tablenotetext{d}{Reference~--~\\ R97~:~\citep{Reynolds1997},~~
Gi00~:~\citep{Gilli2000}, ~T97~:~\citep{Turner1997}, \\
Tu00~:~\citep{Turner2000}, ~Gu03~:~\citep{Guainazzi2003}, ~G98~:~\citep{George1998}, \\
BL02~:~\citep{Blustin2002}, ~Go03(1)~:~\citep{Gondoin2003a}, ~Go03(2)~:~\citep{Gondoin2003b}, \\
Ma02~:~\citep{Malizia2002}, ~T96~:~\citep{Turner1996}, \\
XMM~:~Our own data analysis with {\it XMM-Newton}
observation, ~ASCA~:~Our own data analysis with {\it ASCA} observation}
\chg{\tablenotetext{e}{MOS/PN of {\it XMM-Newton}, SIS/GIS of {\it ASCA} 
or MECS/PDS of {\it Beppo-SAX} observation}}
\tablenotetext{f}{The source was excluded from the result of mean photon index.} 
\tablenotetext{g}{Neutral dual absorber with covering fraction $\sim$90\%. please see Table.2 in \citep{Gondoin2003a}.}
\tablenotetext{(*)}{For these sources, we have used the spectral fit results 
involving models with complicated absorption features. 
See Sect.~\ref{sec:SpecialAGN} for details.}
\end{deluxetable}

%{Results~--~Sample~1 \label{table:3}
\begin{deluxetable}{clccccccc}
\tabletypesize{\scriptsize}
\tablecaption{Results~--~Sample~2 \label{table:2-2}}
\tablewidth{0pt}
\tableheadfrac{0.1}
\tablehead{
\colhead{Source ID} & \colhead{Name} &
\colhead{Flux\tablenotemark{a}}   & \colhead{L$_{\rm X}$\tablenotemark{b}}  &
\colhead{Photon Index} &\colhead{Intrinsic N$_{\rm H}$\tablenotemark{c}} &
\colhead{Flux\tablenotemark{a}} & \colhead{Obs.(ref)\tablenotemark{d}}
& \colhead{Exp.time\tablenotemark{e}}\\
& & \colhead{HEAO-1} & \colhead{(10$^{43}$~erg/s)} & &
\colhead{(10$^{20}$~cm$^{-2}$) } & \colhead{ASCA-XMM}&\colhead{(ks)}}
\startdata
29 & Kaz 102 (*)      & 3.3  &  232.   & 1.9(fixed)\tablenotemark{f}&$\leq 1$            & .43   & ASCA  & 16.8/19.7 \\
30 & KUV 18217+6419   & 2.0  &  594.   & 1.906$^{+0.017}_{-0.017}$&0$^{+0.93}$           & 1.6   & ASCA  & 40.0/43.9 \\
31 & MKN 885          & 2.3  &    3.6  & 1.953$^{+0.069}_{-0.069}$&16.8$^{+4.2}_{-4.2}$  & 0.6   & ASCA  & 18.5/21.5 \\
32 & MKN 876          & 0.97 &   43.   & 1.971$^{+0.035}_{-0.034}$&0$^{+0.27}$           & 0.49  & XMM   & 7.2/2.6   \\
33 & 3C 390.3         & 1.1  &    8.5  & 1.635$^{+0.022}_{-0.021}$&2.9$^{+1.4}_{-1.5}$   & 1.8   & ASCA  & 39.1/46.7 \\
34 & MKN 290          & 1.5  &    3.2  & 1.613$^{+0.032}_{-0.031}$&4.4$^{+2.0}_{-1.9}$   & 1.0   & ASCA  & 40.1/45.5 \\
35 & MKN 279          & 2.5  &    5.0  & 1.764$^{+0.037}_{-0.038}$&0$^{+4.4}$            & 1.7   & XMM   & 29.2/26.3 \\
36 & H 1318+692       & 1.2  &   13.   & 1.823$^{+0.208}_{-0.118}$&2.3$^{+12.5}_{-2.3}$  & .15   & ASCA  & 19.7/22.0 \\
37 & H 1419+480       & 2.0  &   26.   & 1.837$^{+0.033}_{-0.031}$&6.8$^{+1.7}_{-1.7}$   & 0.7   & XMM   & 9.9/7.2   \\
38 & H 1320+551       & 2.1  &   22.   & 1.679$^{+0.254}_{-0.203}$&7.5$^{+15.3}_{-7.5}$  & .10   & ASCA  & 10.1/11.6 \\
39 & PG 0804+761      & 1.1  &   28.   & 2.187$^{+0.027}_{-0.026}$&5.0$^{+1.3}_{-1.4}$   & 1.0   & ASCA  & 41.8/47.6 \\
40 & MKN 506          & 0.91 &    4.0  & 1.932$^{+0.072}_{-0.050}$&1.8$^{+5.0}_{-1.8}$   & .63   & ASCA  & 16.5/18.2 \\
41 & MRK 6(*)         & 3.2  &    3.9  & 1.81(fixed)\tablenotemark{f}&1043               & 1.3   & XMM   & 18.4/26.0 \\
42 & H 1537+339\tablenotemark{g}       & 3.6  &  1350.  & 1.8\tablenotemark{f}&$\leq 1$  & **  &Missing      &       \\
43 & MKN 478          & 0.78 &   11.   & 2.213$^{+0.46}_{-0.12}$  &1.6$^{+126}_{-1.6}$   & .27   & XMM   & 26.2/24.8 \\
44 & MKN 464          & 3.9  &   26.   & 1.590$^{+0.215}_{-0.208}$&66.1$^{+70}_{-66}$    & .27   & XMM   & 7.5/5.5   \\
45 & NGC 5033         & 3.0  &    0.15 & 1.668$^{+0.091}_{-0.087}$&0.27$^{+7.7}_{-0.27}$ & .26   & ASCA  & 35.7/38.8 \\
46 & PG 1425+267      & 1.1  &  595.   & 1.614$^{+0.115}_{-0.110}$&27.4$^{+23.7}_{-22.7}$& .28   & ASCA  & 38.1/41.7 \\
47 & NGC 4151         & 6.9  &    0.53 & 1.93                     &340                   & 4.4   & Sc03  & 110/91    \\
48 & AKN 564          & 0.71 &    0.98 & 2.192$^{+0.051}_{-0.071}$&0$^{+7.9}$            & 1.5   & XMM   & 11.8/0    \\
49 & NGC 5548         & 3.0  &    2.1  & 1.565$^{+0.016}_{-0.017}$&0$^{+3.2}$            & 5.3   & XMM   & 27.3/19.8 \\
\enddata

\tablenotetext{a}{Observed After absorption,~(10$^{-11}$~erg~cm$^{-2}$~s$^{-1}$)}
\tablenotetext{b}{Intrinsic X-ray luminosity in $2-10$~keV band (erg~s$^{-1}$)}
\tablenotetext{c}{After subtracted with Galactic N$_{\rm H}$ }
\tablenotetext{d}{Reference~--~\\
Sc03~:~\citep{Schurch2003}, \\
XMM~:~Our own data analysis with {\it XMM-Newton} observation,
ASCA~:~Our own data analysis with {\it ASCA} observation}
\chg{\tablenotetext{e}{MOS$/$PN of {\it XMM-Newton}, SIS$/$GIS of {\it ASCA}
or MECS$/$PDS of {\it Beppo-SAX} observation}}
\tablenotetext{f}{The source was excluded from the result of mean photon index.} 
\tablenotetext{g}{No data with ASCA or XMM-Newton observation}
\tablenotetext{(*)}{For these sources, we have used the spectral fit results 
involving models with complicated absorption features. 
See Sect.~\ref{sec:SpecialAGN} for details.}
\end{deluxetable}
%Results~--~Sample~2
\begin{deluxetable}{cllllcc}
\tabletypesize{\scriptsize}
\tablecaption{Spectal Results for Soft excess \label{table:3}}
\tablewidth{0pt}
\tableheadfrac{0.1}
\tablehead{
\colhead{Source ID} & \colhead{Name} &
\colhead{Model} & \colhead{$\Gamma$ or Temp.}  &
\colhead{Dual absorber N$_{\rm H}$} &
\colhead{Intrinsic N$_{\rm H}$\tablenotemark{c}} &
\colhead{$\chi ^2/\nu$ matrix} \\
& & & & \colhead{and Covering fraction\tablenotemark{c}} & & \\
& & & & \colhead{(10$^{20}$~cm$^{-2}$)} & \colhead{(10$^{20}$~cm$^{-2}$)} & 
}
\startdata
1  & IC 4329A         &  blackbody + 4 egde   & 188$^{+9}_{-3}$~eV & & & 1704/1387 \\
2  & 3C 273           &  2 power-law          & 3.768$^{+0.195}_{-236}$ & & & 881/817 \\
4  & NGC 5506         &  mekal + dual absorber& 86$^{+10}_{-6}$~eV & 220$^{+12}_{-6}$ (95\%) & 94$^{+11}_{-7}$ & 780/747 \\
9  & MKN 509          &  2 power-law          & 3.412$^{+0.121}_{-0.087}$ & & & 465/427 \\
13 & MKN 926          &  2 power-law          & 2.998$^{+0.654}_{-0.278}$ & & & 251/221 \\
14 & NGC 7469         &  2 power-law          & 4.611$^{+0.221}_{-0.235}$ & & & 655/607 \\
15 & NGC 4593         &  bremss + edge        & 358$^{+26}_{-30}$~eV      & & & 376/303 \\
19 & Fairall 49       &  broken power-law     & 4.646$^{+0.810}_{-0.617}$ & & & 1575/1415 \\
20 & 3C 445(*)        &  dual absorber        & & 1321$^{+114}_{-158}$ (90\%) & & 202/159 \\
22 & NGC 7314         &  blackbody            & 55$^{+2}_{-2}$~eV         & & & 1404/1151 \\
24 & ESO 103-G35      &  blackbody + 2 dual absorber & 77$^{+22}_{-31}$~eV & 1145$^{+357}_{-427}$ (94\%) & 168$^{+140}_{-102}$& 185/157   \\
   &                  &                       &                           & 1032$^{+483}_{-401}$ (95\%)& & \\
26 & MRK 590          &  2 power-law          & 3.194$^{+0.298}_{-0.304}$ & & & 403/365 \\
28 & Fairall 9        &  2 power-law          & 3.892$^{+0.276}_{-0.216}$ & & & 881/789 \\
32 & MKN 876          &  2 power-law          & 5.162$^{+0.353}_{-0.564}$ & & & 202/156\tablenotemark{a} \\
35 & MKN 279          &  bremss + edge        & 406$^{+22}_{-26}$~eV & & & 663/563 \\
37 & H 1419+480       &  blackbody + edge     & 180$^{+22}_{-78}$~eV & & & 700/587 \\
%%41 & MRK 6(*)         &  2 dual absorber + pexrav  & Rel-refl = 1.21\tablenotemark{b} & 211$^{+12}_{-46}$ (91\%) & & 476/422 \\
%%   &                  &                       & cos$i$ = 95\%  & 638$^{+71}_{-24}$ (57\%)  & & \\
43 & MKN 478          &  bremss               & 266$^{+9}_{-13}$~eV & & & 321/275 \\
44 & MKN 464          &  2 power-law + edge   & 5.027$^{+0.486}_{-0.526}$ & & & 263/225 \\
48 & AKN 564          &  2 power-law          & 4.915$^{+0.211}_{-0.244}$ & & & 286/234 \\
49 & NGC 5548         &  blackbody + 3 edge   & 187$^{+1}_{-3}$~eV & & & 842/679 \\
\enddata

\tablenotetext{a}{1.5-3.0~keV band was extracted for spectral fitting.}
\tablenotetext{b}{Reflection fraction}
\tablenotetext{c}{After subtracted with Galactic N$_{\rm H}$ }
\tablenotetext{(*)}{For these sources, we have used the spectral fit results 
involving models with complicated absorption features. 
See Sect.~\ref{sec:SpecialAGN} for details.}
\end{deluxetable}
%{Spectal Results for Soft excess }
\begin{deluxetable}{clccr}
\tabletypesize{\scriptsize}
\tablecaption{Results of Iron Line \label{table:4}}
\tablewidth{0pt}
\tableheadfrac{0.1}
\tablehead{
\colhead{Source ID} & \colhead{Name} & \colhead{line energy} 
& \colhead{normalization} & \colhead{Equivalent width } \\
& & \colhead{(keV)} & (MOS or GIS) & (MOS or GIS)
}
\startdata
1  & IC 4329A         &  6.382$^{+0.031}_{-0.031}$ & 1.46$^{+0.45}_{-0.41}$e-4 & 76 eV \\
4  & NGC 5506         &  6.435$^{+0.053}_{-0.050}$ & 7.29$^{+3.96}_{-5.60}$e-5 & 52 eV \\
9  & MKN 509          &  6.302$^{+0.202}_{-0.206}$ & 4.96$^{+5.59}_{-1.93}$e-5 & 125 eV \\
14 & NGC 7469         &  6.458$^{+0.067}_{-0.286}$ & 2.96$^{+2.41}_{-1.63}$e-5 & 103 eV \\
15 & NGC 4593         &  6.296$^{+0.075}_{-0.091}$ & 5.96$^{+2.74}_{-2.87}$e-5 & 138 eV \\
19 & Fairall 49       &  6.497$^{+0.138}_{-0.148}$ & 2.69$^{+0.83}_{-0.76}$e-5 & 132 eV \\
20 & 3C 445(*)        &  6.375$^{+0.068}_{-0.240}$ & 2.56$^{+3.21}_{-0.92}$e-5 & 178 eV \\
21 & MKN 1152         &  6.508$^{+0.179}_{-0.097}$ & 1.37$^{+0.75}_{-0.56}$e-5 & 202 eV \\
22 & NGC 7314         &  6.391$^{+0.074}_{-0.046}$ & 4.49$^{+2.18}_{-1.29}$e-5 & 98 eV \\
26 & MRK 590          &  6.369$^{+0.060}_{-0.049}$ & 1.55$^{+822}_{-1.34}$e-5  & 253 eV \\
28 & Fairall 9        &  6.434$^{+0.144}_{-1.983}$ & 9.69$^{+64.6}_{-2.06}$e-5 & 605 eV \\
31 & MKN 885          &  6.056$^{+0.315}_{-0.248}$ & 2.67$^{+1.42}_{-1.66}$e-5 & 418~eV\\
33 & 3C 390.3         &  6.451$^{+0.103}_{-0.103}$ & 4.60$^{+1.20}_{-1.65}$e-5 & 215~eV\\
34 & MKN 290          &  6.484$^{+0.158}_{-0.198}$ & 3.13$^{+822}_{-1.07}$e-5  & 273~eV\\
35 & MKN 279          &  6.480$^{+0.065}_{-0.060}$ & 1.63$^{+1.26}_{-1.23}$e-5 & 86~eV\\
37 & H 1419+480       &  6.431$^{+0.029}_{-0.052}$ & 8.65$^{+8220}_{-3.67}$e-6 & 109~eV\\
%%41 & MRK 6(*)         & 6.434$^{+0.052}_{-0.009}$ & 1.35$^{+822}_{-0.32}$e-5 & 61~eV\\
43 & MKN 478          &  6.661$^{+0.628}_{-0.462}$ & 1.32$^{+1.07}_{-0.75}$e-5 & 506~eV\\
48 & AKN 564          &  6.405$^{+0.434}_{-0.405}$ & 1.71$^{+41.3}_{-0.54}$e-4 & 1400~eV\\
49 & NGC 5548         &  6.407$^{+0.050}_{-0.049}$ & 2.82$^{+1.62}_{-1.56}$e-5 & 26~eV\\
\enddata

\tablenotetext{(*)}{For these sources, we have used the spectral fit results 
involving models with complicated absorption features. 
See Sect.~\ref{sec:SpecialAGN} for details.}
\end{deluxetable}

%{Results of Iron Line \label{table:6}}
\begin{deluxetable}{lcccc}
\tabletypesize{\scriptsize}
\tablecaption{The Best Fit Parameters of the HXLF Models\label{table:5}}
\tablewidth{0pt}
\tableheadfrac{0.1}
\tablehead{
\colhead{Sample} & \colhead{$A$\tablenotemark{a}} & 
\colhead{$\log\,L_*$\tablenotemark{b}} & \colhead{$\gamma_1$} & 
\colhead{$\gamma_2$}
}
\startdata
All\tablenotemark{c}  & 3.68$\pm 0.55$ & 44.02$^{+0.34}_{-0.43}$ & 0.93$^{+0.20}_{-0.29}$ & 2.51$^{+0.53}_{-0.37}$ \\
Unabsorbed  & 1.98$\pm 0.36$ & 44.00(fix) & 0.72$^{+0.17}_{-0.19}$ & 2.34$^{+0.24}_{-0.22}$ \\
Absorbed  & 1.60$\pm 0.41$ & 44.00(fix) & 1.12$^{+0.17}_{-0.19}$ & 3.34$^{+0.90}_{-0.65}$ \\
\hline
U03  & 2.64$\pm 0.18$ & 44.11$\pm 0.23$ & 0.93$\pm 0.13$ & 2.23$\pm 0.15$ \\
\enddata

\tablecomments{Errors are 1 $\sigma$ for a single parameter.}
\tablenotetext{a}{In units of $10^{-6} h^3_{70}$~Mpc$^{-3}$.}
\tablenotetext{b}{In units of $h^{-2}_{70}$~erg s$^{-1}$.}
\tablenotetext{c}{only $z < 0.2$ and $\log\,N_{\rm H} < 24$ sources.}

\end{deluxetable}

%{The Best Fit Parameters of the HXLF Models \label{table:5}}

\end{document}